\documentclass{aa}
\usepackage{mathabx}
\usepackage[title]{appendix}
\usepackage{natbib}
\usepackage{float}
\usepackage{color}
\usepackage{hyperref}
\usepackage{academicons}
\usepackage{xcolor}
\newcommand{\orcid}[1]{\href{https://orcid.org/#1}{\textcolor[HTML]{A6CE39}{\aiOrcid}}}
\usepackage{orcidlink}

\usepackage{graphicx}
\usepackage{txfonts}
\begin{document}

   \title{The SPIRou Legacy Survey}
   \subtitle{{Near-infrared and optical radial velocity analysis of} Gl\,480 and Gl\,382 using SPIRou, HARPS and CARMENES spectrographs}
   \titlerunning{{nIR and optical RV analysis of} Gl\,480 and Gl\,382}

   \author {M. Ould-Elhkim \inst{1}         
          \fnmsep\thanks{Based on observations obtained at the Canada-France-Hawaii Telescope (CFHT) which is operated from the summit of Maunakea by the National Research Council of Canada, the Institut National des Sciences de l'Univers of the Centre National de la Recherche Scientifique of France, and the University of Hawaii. The observations at the Canada-France-Hawaii Telescope were performed with care and respect from the summit of Maunakea which is a significant cultural and historic site. Based on observations obtained with SPIRou, an international project led by Institut de Recherche en Astrophysique et Plan\'etologie, Toulouse, France.
          SPIRou is an acronym for SPectropolarimetre InfraROUge (infrared spectropolarimeter).}
          \orcidlink{0000-0002-1059-0193}
          \and
          C. Moutou\inst{1} \orcidlink{0000-0002-2842-3924}
          \and
          J-F. Donati\inst{1}\orcidlink{0000-0001-5541-2887}
          \and
          \'E. Artigau\inst{2,3}
          \and
          C. Cadieux\inst{2}
        \and
        E. Martioli\inst{4, 5}\orcidlink{0000-0002-5084-168X}
          \and
          T. Forveille\inst{6}
          \and 
          J. Gomes da Silva \inst{7}
            \and
            R. Cloutier\inst{8, 9}\orcidlink{0000-0001-5383-9393}
          \and
          A. Carmona\inst{6}
          \and
          P. Fouqué\inst{1}
         \and
           P. Charpentier\inst{1}
           \and
           P. Larue \inst{6}
          \and
          N.J. Cook\inst{2}
          \and
          X. Delfosse\inst{6}
          \and
          R. Doyon\inst{2}
          }

    \institute{Universit\'e de Toulouse, UPS-OMP, IRAP, 14 avenue E. Belin, Toulouse, F-31400, France, France\\
    \email{merwan.ould-elhkim@irap.omp.eu}     
    \and  
    Trottier Institute for Research on Exoplanets, Université de Montréal, Département de Physique, C.P. 6128 Succ. Centre-ville, Montréal,
    QC H3C 3J7, Canada
    \and 
    Observatoire du Mont-M\'egantic, Universit\'e de Montr\'eal, D\'epartement de Physique, C.P. 6128 Succ. Centre-ville, Montr\'eal, QC H3C 3J7, Canada
    \and
    Laborat\'{o}rio Nacional de Astrof\'{i}sica, Rua Estados
Unidos 154, 37504-364, Itajub\'{a} - MG, Brazil 
    \and
 Institut d'Astrophysique de Paris, CNRS, UMR 7095, Sorbonne
Universit\'{e}, 98 bis bd Arago, 75014 Paris, France
    \and
      Univ. Grenoble Alpes, CNRS, IPAG, 38000 Grenoble, France
    \and
    Instituto de Astrofísica e Ciências do Espaço, Universidade do Porto, CAUP, Rua das Estrelas, 4150-762 Porto, Portugal
    \and 
    Center for Astrophysics | Harvard \& Smithsonian, 60 Garden Street, Cambridge, MA 02138, USA
    \and 
    Department of Physics \& Astronomy, McMaster University, 1280 Main Street West, Hamilton, ON L8S 4L8, Canada}
% \abstract{}{}{}{}{} 
% 5 {} token are mandatory
 
  \abstract
  % context heading (optional)
  % {} leave it empty if necessary  
   {Advancements in the field of exoplanetary research have extended radial velocity (RV) observations from the optical to the near-infrared (nIR) domain. M dwarf stars, characterized by their lower masses and higher prevalence of rocky planets, have become a focal point of investigation. This study uses data from the near-infrared spectropolarimeter SPIRou and data available in the literature from the HARPS and CARMENES spectrographs operating in the optical to analyze RVs of two nearby M dwarfs, Gl\,480 and Gl\,382. }
  % aims heading (mandatory)
   {{This work aims to detect and characterize exoplanetary companions around Gl\,480 and Gl\,382 by mitigating stellar activity effects through advanced data analysis techniques. The study seeks to improve the reliability of RV signals by integrating multi-wavelength observations and stellar activity diagnostics.}}
  % methods heading (mandatory)
  {{The study employs a comprehensive approach that combines the line-by-line (LBL) framework with the \texttt{Wapiti} (Weighted principAl comPonent analysIs reconsTructIon) method to correct for systematics in SPIRou data. Through an extensive analysis of available stellar activity indicators and by combining optical data from the HARPS and CARMENES instruments, we perform a joint analysis of RV measurements in both the nIR and optical domains.}}
  % results heading (mandatory)
 {{Our analysis confirms the detection of a planet orbiting Gl\,480 with a period of $9.5537 \pm 0.0005$,d and a minimum mass of $8.8 \pm 0.7$,M$_\oplus$. Additionally, we detect a tentative signal at 6.4,d, whose significance depends strongly on the choice of Gaussian Process priors constrained by stellar activity indicators and would require further observations for confirmation. In contrast, no planetary signals are detected for Gl\,382, where RV variations are dominated by stellar activity.}}
  % conclusions heading (optional), leave it empty if necessary 
   {}


   \maketitle
%
%-------------------------------------------------------------------

\section{Introduction}

The first detection of an exoplanet around 51\,Pegasi was made possible by the radial velocity (RV) method \citep{1995Natur.378..355M} using the optical spectrograph ELODIE at the Haute-Provence Observatory in France \citep{1996A&AS..119..373B}. Since then, advancements in optical instrumentation with instruments such as HARPS \citep{2003Msngr.114...20M} and ESPRESSO \citep{2014SPIE.9147E..1HM} have resulted in the identification of numerous exoplanetary systems (\citet{Faria_2022, Su_rez_Mascare_o_2023}; to quote just a few of the latest results). Despite having historically started in the optical domain, RV observations have recently extended to the near-infrared (nIR) domain with instruments such as GIANO \citep{2016ExA....41..351C}, HPF \citep{Metcalf2019}, CARMENES-NIR \citep{bauer2020}, IRD \citep{hirano2020}, SPIRou \citep{donati2020} and NIRPS \citep{2022SPIE12184E..54T}. The nIR does not benefit from the extensive research that has been conducted in the optical domain with respect to instrumental and data-reduction techniques. Most notably, the nIR comes with new challenges such as telluric correction, instrument stability at cryogenic temperatures, wavelength calibration, as well as persistence and other yet to be investigated effects of infrared arrays \citep{2018SPIE10709E..1PA, 2022ascl.soft11019C}. However, this spectral range provides a unique opportunity to explore the $YJHK$ spectral bandpasses \citep{quirrenbach_carmenes_2018, 2020ApJS..247...11R}, which are especially rich in molecular bands such as CO, OH, and H$_2$O, in addition to being the domain in which cool stars are the most luminous.

\par

The nIR is thus of interest to better detect and characterize exoplanetary systems around M dwarfs \citep{2022AJ....164...96C,2022A&A...660A..86M, 2023arXiv230801454A}. Those stars have attracted a great deal of interest due to their abundance in the solar neighborhood \citep{2006AJ....132.2360H}. They are of particular significance as they have lower masses, making them excellent candidates for detecting {small mass exoplanets since it increases their RV signatures in the  m\,.s$^{-1}$ range} \citep{2016Natur.536..437A, 2013A&A...549A.109B, Faria_2022, Su_rez_Mascare_o_2023}. Additionally, these stars are known to host a larger number of rocky planets than other star types \citep{Bonfils_2013, DressingCharbonneau2015, 2016MNRAS.457.2877G, 2020MNRAS.498.2249H, 2021A&A...653A.114S}. 

\par

On the other hand, M dwarfs are known to exhibit a high level of magnetic activity, which can lead to spurious RV signals \citep{2007A&A...474..293B, Gomes_da_Silva_2012, 2016ApJ...821L..19N, 2016MNRAS.461.1465H}. However, this activity is wavelength-dependent \citep{reiners2013, baroch2020} and less important in the nIR. Consequently, an example of the potential synergy between both optical and nIR is to conduct RV measurements in both domains so as to confirm or rule out the planetary nature of ambiguous signals \citep{pia2023, carmona2023nearir}.

\par

Within this context, the SPIRou Legacy Survey Planet Search program (SLS-PS\footnote{https://spirou-legacy.irap.omp.eu/doku.php}) is a large program that aims at detecting and characterizing planetary systems around approximately $\sim$50 nearby M dwarfs \citep{2023arXiv230711569M}. The program was allocated 153 nights of observation time at the Canada-France-Hawaii Telescope (CFHT) from February 2019 to June 2022. In this paper, we present the analysis of the nearby M dwarfs Gl\,480 and Gl\,382 using SPIRou data obtained as part of this program and other publicly available datasets. 

\par 

This paper is organized as follows. In Section \ref{sec:observation}, we describe our targets and the RV data used in our analysis as well as an overview of the Wapiti correction method to mitigate systematics in the SPIRou data of our two targets. Section \ref{sec:analysis} presents a separate analysis for each target, discussing their characteristics and the determination of orbital parameters and we conclude in Section \ref{sec:conclusion}.

\section{Observations}\label{sec:observation}

\subsection{The SPIRou spectropolarimeter, the \texttt{APERO} reduction and the LBL algorithm}

SPIRou is a high-resolution near-infrared spectropolarimeter installed in 2018 at the Cassegrain focus of CFHT. It has a spectral range going from 0.98 to 2.35 $\mu$m at a resolving power of 70,000, and is stabilized in a vacuum chamber with milli-K temperature stability \citep{2018SPIE10702E..62C, donati2020}. A detailed presentation of the instrumental design and early measurement performances is provided in \citet{donati2020}.

\par 

All SPIRou spectra are reduced using the {0.7.288} version of the SPIRou data reduction system (DRS), called \texttt{APERO} (A PipelinE to Reduce Observations) detailed in \citet{2022ascl.soft11019C}. \texttt{APERO} processes science frames to produce 2D and 1D spectra from the two science channels and the RV reference channel, and it uses a combination of exposures from a UNe hollow cathode lamp and a Fabry-Pérot etalon for wavelength calibration \citep{2021A&A...648A..48H}. Barycentric Earth radial velocity (BERV) is calculated using \textsc{barycorrpy} \citep{2018RNAAS...2....4K} to shift the spectra to the barycentric frame of the solar system. \texttt{APERO} also performs optimal extraction of both science channels and the Fabry-Pérot RV reference channel, flat-fielding, a correction of thermal background, and a correction of the residual leak of the Fabry-Pérot RV reference spectrum onto the science channel. Based on the TAPAS model spectrum of the Earth atmosphere \citep{2014hitr.confE...8B} and a library of hot-star spectra obtained with SPIRou in various conditions of airmass and water vapor, APERO corrects the science spectra from telluric spectrum using a three-step telluric correction approach which will be detailed in an upcoming publication (Artigau et al. (2025), in prep.). The final product is a wavelenght calibrated telluric corrected 2D spectrum

\par

RVs are subsequently computed from the telluric-corrected spectra using the {0.63.007 version} of the Line-By-Line (LBL\footnote{https://lbl.exoplanets.ca/}) method. This method is based on the framework developed in \citet{2001A&A...374..733B} and was first proposed by \citet{Dumusque_2018} then further explored by \citet{Cretignier_2020} for optical RV observations with HARPS and more recently revisited by \citet{2022AJ....164...84A} in the context of nIR observations. The LBL algorithm requires a high-signal-to-noise ratio (S/N) template spectrum of the observed star, simply called template in the following, and derived from the median of all collected spectra. RVs are then computed by comparing each spectrum to the template and its first few derivatives. The LBL algorithm is applied to both the stellar extracted spectrum and the Fabry-Pérot spectrum in the RV reference channel, with the latter providing a precise measurement of the instrumental drift at the exact time of our observations. The final RV time series is corrected from this drift {as well as from a nightly zero-point (NZP) correction in the time domain to eliminate long-term instrumental variations. The list of standard stars used to compute this NZP is GJ\,905, GJ\,699, GJ\,1012, GJ\,1289, GJ\,3378, GJ\,752\,A, GJ\,1002, GJ\,1286, GJ\,1103 and PMJ21463+3813 (same \texttt{APERO} version)}.

\subsection{Targets}

Gl\,480 (Wolf\,433) is a red dwarf star with spectral type M3.5V \citep{2013AJ....145..102L} that is located $14.2625 \pm 0.0060$ pc away from Earth \citep{2020yCat.1350....0G}. In \citet{2022MNRAS.516.3802C}, its physical properties were determined from the analysis of SPIRou spectra and are summarized in Table \ref{stellar_properties}. Additionally, time series of the longitudinal magnetic field ($B_\ell$) of this star were obtained from circularly polarized spectropolarimetric sequences. However, no clear periodic variation was detected in the time series \citep{2023A&A...672A..52F}. In the study by \citet{2023arXiv230714190D}, which is a parallel analysis of the $B_\ell$ time series using different data reduction and analysis of the spectra, the authors identified a rotational period $P_{rot}$ of $25.00 \pm 0.36$\,d. However, a prior study by \citet{2020ApJS..250...29F} reported an activity-related signal occurring at a period of $49.3 \pm 0.2$\,d. It is plausible that the 25\,d signal observed in the $B_\ell$ data might actually represent the first harmonic of the true rotational period. Finally, a planet orbiting at a $9.567 \pm 0.005$\,d orbital period was claimed to orbit Gl\,480 from the analysis of HIRES and HARPS data \citep{2020ApJS..250...29F}. The signal had a semi-amplitude $K=6.80 \pm 0.87$\,m\,s$^{-1}$ with an estimated $Mp \sin i$ of $13.2 \pm 1.7\ M_\Earth$.

\par 

Gl\,382  (V* AN Sex) is a red dwarf of spectral type M2.0V \citep{1991ApJS...77..417K} located $7.707 \pm 0.0015$ pc away from Eath \citep{2020yCat.1350....0G}. Its physical properties are also listed in  Table \ref{stellar_properties}. time series of $B_\ell$ of this star were also obtained from Stokes V spectropolarimetric sequences and led to the detection and characterization of a signal at $21.32^{+0.04}_{-0.03}$\,d \citep{2023A&A...672A..52F} ($21.91 \pm 0.16$\,d in \citet{2023arXiv230714190D}). {Its rotational period is rather well-established in the literature using other instruments such as ASAS photometric data ($21.6 \pm 2.2$\,d; \citet{kiraga2011agerotationactivity}) or HARPS ($21.7 \pm 0.1$\,d; \citet{2015MNRAS.452.2745S})}. In \citet{2021A&A...653A.114S}, a signal with a period of 10.65\,d was detected in CARMENES optical RV data. However, the signal was automatically flagged as spurious due to this period being close to the first harmonic of the rotational period. This star was also part of the recent activity study conducted by \citet{2023arXiv230303998M} from HARPS spectra, and they did not find any signs of an activity cycle in the chromospheric activity indices {even though a previous study had detected such cycle with a period of $13.6 \pm 1.7$ years at a 3$\sigma$ level \citep{2016A&A...595A..12S}}.

\begin{table*}[h!]
\centering
\caption[]{Stellar parameters of Gl\,480 and Gl\,382.}
\begin{tabular}{c c c c}
\hline
\noalign{\smallskip}
Parameters & Gl\,480 & Gl\,382 & Ref. \\
\noalign{\smallskip}
\hline
\noalign{\smallskip}
\multicolumn{4}{c}{\textbf{Stellar parameters}} \\
\noalign{\smallskip}
\hline
\noalign{\smallskip}
$\mathrm{T_{eff}}$ (K)& $3509 \pm 31$ & $3644 \pm 31$ & 1\\
$\left[M/H\right]$ & $0.26 \pm 0.10$ & $0.15 \pm 0.10$ & 1\\
$\left[\alpha/\texttt{Fe}\right]$ & $-0.01 \pm 0.04$ & $-0.02 \pm 0.04$ & 1\\
$\log\ g$ & $4.88 \pm 0.06$ & $4.75 \pm 0.05$ & 1\\
Radius $\mathrm{(R_\odot)}$ & $0.449 \pm 0.008$ & $0.511 \pm 0.009$ & 1\\
Mass $\mathrm{(M_\odot)}$ & $0.45 \pm 0.02$ & $0.51 \pm 0.02$ & 1\\
$\mathrm{\log \ (L/L_\odot)}$ & $-1.562 \pm 0.002$ & $-1.384 \pm 0.001$ & 1\\
$\mathrm{mean\ B_\ell}$ (G) & $6.6^{+3.5}_{-3.3}$ & $-0.7^{+4.2}_{-4.2}$ & 2\\
$\mathrm{P_{rot}}$ (d) & - &  $21.32^{+0.04}_{-0.03}$ & 2\\
$\mathrm{P_{rot}}$ (d) & $25.00 \pm 0.24$ &  $21.91 \pm 0.16$ & 3\\
$\mathrm{P_{rot}}$ (d) & $49.3 \pm 0.2$ & - & 4\\
$\mathrm{P_{rot}}$ (d) &$50.3^{+0.9}_{-0.9}$ & $21.41^{+0.14}_{-0.14}$ & This work \\
\hline
\noalign{\smallskip}
\multicolumn{4}{c}{\textbf{Astrometry}} \\
\noalign{\smallskip}
\hline
\noalign{\smallskip}

Right ascension (J2000) & $12^h38^m52.44^s$ &  $10^h12^m17.67^s$ & 5 \\
\noalign{\smallskip}
Declination (J2000) & $+11^\degree 41'46.14"$ & $-03^\degree 44'44.39"$ & 5\\
\noalign{\smallskip}
Parallax (mas) & $70.1139 \pm 0.0297$ & $129.7544 \pm 0.0252$ & 5 \\
\noalign{\smallskip}
Distance (pc) & $14.2625 \pm 0.0060$ & $7.707 \pm 0.0015$ & 5 \\
\noalign{\smallskip}
Proper motion RA (mas/year) & $-1155.756 \pm 0.034$ & $-152.760 \pm 0.028$ & 5 \\
\noalign{\smallskip}
Proper motion D (mas/year) & $-250.862 \pm 0.038$ & $-243.693 \pm 0.028$ & 5 \\
\noalign{\smallskip}
\hline
\noalign{\smallskip}
\multicolumn{4}{c}{\textbf{Photometry}} \\
\noalign{\smallskip}
\hline
\noalign{\smallskip}
G [mag] & $10.335732 \pm 0.002806$ & $8.331761 \pm 0.002821$ & 5 \\
\hline
\noalign{\smallskip}
\end{tabular}
\begin{minipage}{\textwidth} % This ensures the note is centered and spans the table width
\small
\textbf{1}: \citet{2022MNRAS.516.3802C}, \textbf{2}: \citet{2023A&A...672A..52F}, \textbf{3}: \citet{2023arXiv230714190D}, \textbf{4}: \citet{2020ApJS..250...29F}, \textbf{5}: \citet{2020yCat.1350....0G}.
\end{minipage}
\label{stellar_properties}

\end{table*}

\subsection{RVs data sets}

\subsubsection{Gl\,480}

 SPIRou observations for Gl\,480 spans from November 13, 2019 to  {June 20, 2024}, {and consists in 130 data points from the SLS program and other programs (PI. Carmona and PI. Artigau)\footnote{CFHT program: P45. From semester 19A to 24B included.}}. The airmass during these observations has a median value of 1.3, the signal-to-noise ratio per 2.28 km\,s$^{-1}$ pixel bin in the middle of the $H$ band has a median value of {110.7} and the median of the error bars is {1.7\,m\,s$^{-1}$}.

\par 

We also retrieved RV time series collected with HARPS and CARMENES (optical data). {We found 37 RV measurements from HARPS in \citet{2020A&A...636A..74T} (RMS $= 6.1$\,m\,s$^{-1}$, error bars median value of 1.5\,m\,s$^{-1}$)) and 7 from CARMENES in \citet{Ribas_2023} (RMS $= 3.3$\,m\,s$^{-1}$, error bars median value of 1.2\,m\,s$^{-1}$)), which were both reduced using the SERVAL pipeline \citep{2018A&A...609A..12Z}.}

\subsubsection{Gl\,382}

SPIRou observations were again conducted using circular polarization sequences. Observations for Gl\,382 were carried out from February 02, 2019 to June 07, 2023, {and consists in 146 data points from the SLS program and other programs (PI. Carmona and PI. Artigau)\footnote{CFHT program: Q57. From semester 19A to 23B included.}}. The median airmass during these observations was {1.4}, the median signal-to-noise ratio per 2.28 km\,s$^{-1}$ pixel bin in the middle of the $H$ band was 154.5{and the median of the error bars is 1.6\,m\,s$^{-1}$}.

\par 

In addition to the SPIRou data, RV time series were {also} retrieved from HARPS, and CARMENES (optical data). We used 32 HARPS (RMS $= 6.7$\,m\,s$^{-1}$, error bars median value of 1.0\,m\,s$^{-1}$)) and 77 CARMENES (RMS $= 5.5$\,m\,s$^{-1}$, error bars median value of 1.5\,m\,s$^{-1}$)) RV measurements from \citet{2020A&A...636A..74T} and \citet{Ribas_2023}, respectively. These measurements were reduced using the SERVAL pipeline \citep{2018A&A...609A..12Z}.

\subsection{\texttt{Wapiti} correction}\label{sec:method}

The LBL method provides a mean RV value of the stellar spectrum and per-line RV time series on which it is possible to apply a weighted Principal Component Analysis (wPCA\footnote{https://github.com/jakevdp/wpca}; \citet{{2015MNRAS.446.3545D}}). The \texttt{Wapiti}\footnote{https://github.com/HkmMerwan/wapiti} method \citep{2023arXiv230502123O} uses this approach and applies a wPCA to the LBL per-line RV time series {in order to find proxy time series of any remaining} systematic of various origins. In summary, it uses a wPCA in the per-line RV time series (corrected for drift and NZP) and automatically selects an appropriate number of components through an algorithm slightly different from the one outlined in \citet{2023arXiv230502123O}. In summary, we retained the permutation test described in this work, {with the key difference that, instead of evaluating the significance of the explained variance of a component, we assessed how well it fit the RV. Specifically, we calculated the Bayesian Information Criterion (BIC; \citet{bic}) for each component when fitted individually to the RV (including an offset). We then compared this to the BIC$_0$ value of a model containing only the offset. To assess the robustness of the fit, we tested the log Bayes Factor between these two models, defined as}:

\begin{equation}\label{eq:model_0}
  \log BF = \frac{BIC_0 - BIC}{2}
\end{equation}

{Another modification to the original version of \texttt{Wapiti} is that we replaced the leave-p-out cross-validation by a simple reordering of the principal components once they have been selected. Specifically, we rearrange the components in descending order of their Bayesian Information Criterion (BIC) values. During this process, we identify the component that, when included, minimizes the BIC and designate it as the new primary component. We then iteratively determine subsequent components: for each step, we add the next component that, when combined with the already selected components, further minimizes the BIC. This procedure produces a reordered sequence of principal vectors, denoted as $\hat{V_i}$.} Using the mathematical framework established in \citet{yarara2}, this approach corresponds to calculating the BIC when fitting the following model to the data for the k-th component:

\begin{equation}\label{eq:model_0}
   RV(t) = \gamma_{SPIRou} + \sum_{i=1}^{N}a_i V_i.
\end{equation}

{Where we included an offset term $\gamma_{SPIRou}$ and the coefficients $a_i$ to fit the principal vectors $V_i$ in the model. The parameter $N$ represents the number of components used in the reconstruction, and it can be set to 0, corresponding to the data without applying \texttt{Wapiti} correction. We fitted these models through weighted least-squares regression using the \texttt{statsmodels} module \citep{seabold2010statsmodels}, which also provided their BIC value.}

\par 

{To determine the final number of selected components, we simply choose the number of components that results in the lowest BIC value. This reordering method is similar to the $\chi^2$ reordering technique employed in \citet{cameronpca}, with the only difference being our use of the BIC as the criterion to penalize using too many components.}

\par 

{We can detect the presence of periodic signals in the time series through a Bayesian periodogram \citep{Delisle2018} that compares two models, one model $\mu_0$ corresponding to equation \ref{eq:model_0} and one model $\mu_P$ for each period $P$ defined as}

\begin{equation}\label{eq:model_p}
    RV(t) = \gamma_{SPIRou} + \sum_{i=1}^{N}a_i \hat{V_i} + A \sin\left(2\pi t / P\right) + B \sin\left(2\pi t / P\right),
\end{equation}

{and by computing their respective BIC ($BIC_0$ and $BIC(P)$) and the logarithm of the Bayes Factor $\log BF(P) = \frac{BIC_0 - BIC(P)}{2}$. We computed this periodogram  with a grid of 10,000 points ranging from 1.5\,d to 10,000\,d with a logarithmic spacing, and a value above 5 was used as a threshold to consider a signal statistically significant \citep{Delisle2018}.}

\par 

{In the case of Gl\,480, {four outliers were removed from the dataset}, prior to employing the \texttt{Wapiti} method. Only the fourth wPCA component was found appropriate for this target and the application of the BIC re-ordering was thus not necessary.} 

\par 

Regarding Gl\,382, no evident outliers were detected in the RV time series. {The components 1, 2, 5, 6 and 7 were found appropriate for this target and the application of the BIC re-ordering led to the use of 4 components (In order $V_6$, $V_{2}$, $V_{5}$ and $V_{7}$).}

\par 

In the following section, a comprehensive analysis of Gl\,480 and Gl\,382 will be presented, using the {SPIRou} RV time series along with other available datasets. The analysis will be conducted separately for each target. {Firstly, we will begin by characterizing the stellar activity of the observed targets. Secondly, we will perform an in-depth RV analysis of the targets over different combinations of the datasets and accurately determine the orbital parameters associated with their planetary companion, if any.}

\section{Analysis and results}\label{sec:analysis}

\subsection{Gl\,480}

\subsubsection{Stellar activity}

\begin{figure*}
    \centering
    \includegraphics[width=\linewidth]{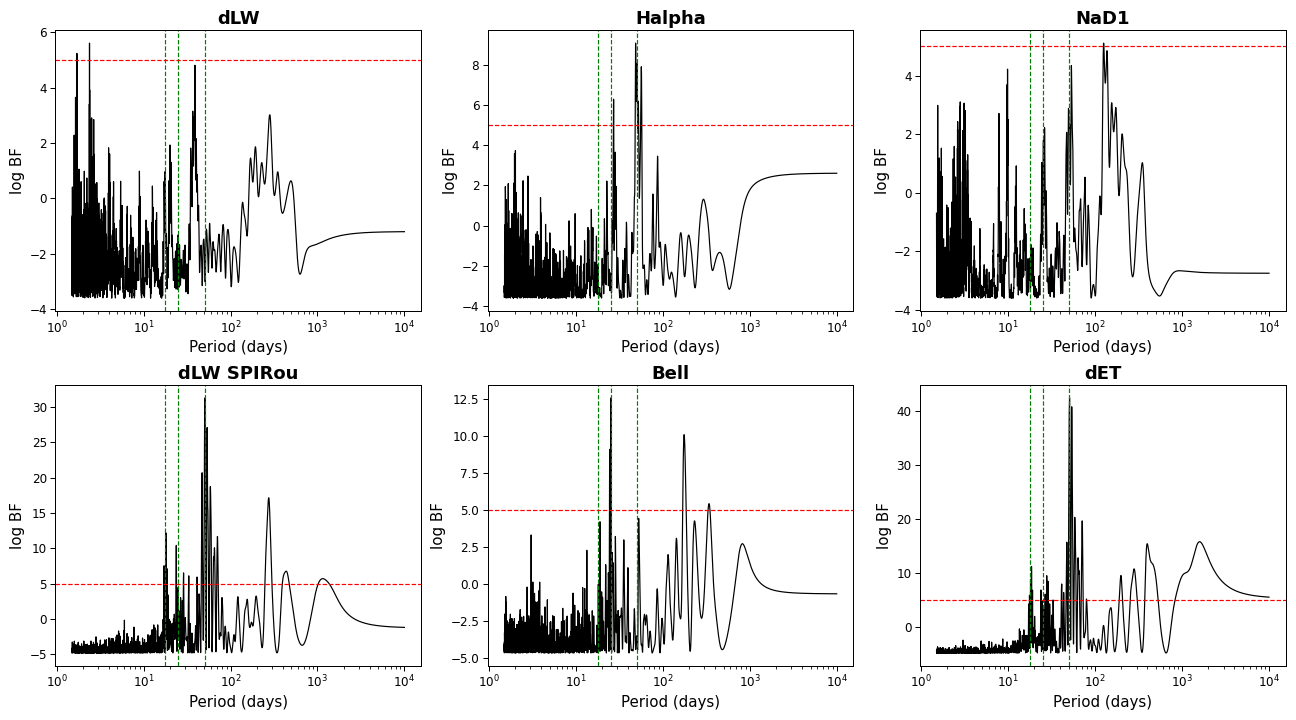}
    \caption{Bayesian periodograms of various stellar activity indicators of Gl\,480. In order dLW, $H\alpha$ and the NaD1 index time series from HARPS data, and the dLW, B$\ell$ and dET time series from SPIRou data. The dashed green lines in the bottom panel mark periodicities at the stellar rotation period (P$_{\mathbf{rot}}$), half rotation period (P$_{\mathbf{rot}}$/2), and one-third rotation period (P$_{\mathbf{rot}}$/3).  The significance level of log BF $= 5$ is indicated by the horizontal red dashed line.}
    \label{fig:gl480_activity}
\end{figure*}

In their study, \citet{2020ApJS..250...29F} performed an activity analysis that detected a signal with a period of $49.3 \pm 0.2$\,d in two stellar indicators, $H\alpha$ and {NaD1}, in the HARPS data. We thus examined those indicators as well as the differential line width (dLW), which is another product obtained from SERVAL \citep{serval}, in the HARPS data. We analyzed the presence of the signals in those indicators {using Bayesian periodograms} and we found no evidence of activity at the reported period in dLW, while we did observe the reported signal in {NaD1} and $H\alpha$ {even though not at a significant level in the former} (see Figure \ref{fig:gl480_activity}).

\par

Regarding SPIRou, {we used three stellar activity indicators: dLW, $B_\ell$ and a new proxy for activity, the differential effective temperature (dET; \citet{dtemp}). This stellar activity proxy is obtained with the LBL analysis from the temperature variation of the star derived from the modulation of spectral line depths in SPIRou spectra. The periodograms can be seen in Figure \ref{fig:gl480_activity}, we found a strong signal at the reported period of \citet{2020ApJS..250...29F} in dLW and dET while $B_\ell$ appears to have a strong signal around its first harmonic.}

\par

{Since the rotational period is the strongest in the dET time series, we decided to use it to obtain a characterization of the rotational period by applying a quasi-periodic Gaussian process (GP) with the following form}:

\begin{equation}
    k\left(\tau\right) = A \exp\left(- \frac{\tau^2}{2\lambda^2} - \Gamma \sin^2\left(\frac{\pi}{P}\tau \right)\right).
    \label{gp_form}
\end{equation}

{The GP was computed using the \texttt{george} package \citep{2015ITPAM..38..252A} and to estimate the hyper-parameters we employed a MCMC approach with the \texttt{emcee} module \citep{2013ascl.soft03002F} by generating 10,000 samples with a burn-in of {2,500} samples and 100 walkers. We obtained a value for the rotational period of {$P_{rot} = 50.6^{+0.8}_{-0.8}$\,d}, consistent with the value of \citet{2020ApJS..250...29F}. The corner plot of the hyper-parameters can be seen in the Appendix \ref{sec:RVanalysis480}}

\subsubsection{SPIRou RV analysis}

{To search for signals in the SPIRou data of in Gl\,480 we adopted an iterative approach using Bayesian periodograms where we systematically add significant signals in the model and then recompute a log BF periodogram. We continue this process while the maximum peak in the periodogram remains above 5 \citep{Delisle2018}. To be more precise, {with $N$ the number of Wapiti systematics} used and $N_p$ representing the number of fitted planets (which can be 0) of periods $P_i$, we compare the models:}

\begin{equation}\label{model_joint}
\begin{aligned}
    M_{0} &= \gamma + \sum_{i=1}^{N}a_i \hat{V_i} + \sum_{i=1}^{N_p} A_i \sin\left(2\pi t / P_i\right) + B_i \sin\left(2\pi t / P_i\right)\\
    M_{P} &=  M_{0} + A \sin\left(2\pi t / P\right) + B \sin\left(2\pi t / P\right). \\
\end{aligned}
\end{equation}

{In this equation, $M_0$ represents the base model, which includes a constant term $\gamma$ diferent for all instruments.}

\par

{If at the $N_p + 1$ iteration a signal at period $P$ is found significant, a circular model $\mu_C$ of the form:}

\begin{align}
\mu_C &= \gamma + \sum_{i=1}^{N}a_i \hat{V_i} 
+ \sum_{i=1}^{N_p} K_i \sin\left(2\pi \frac{t - T_i}{P_i} \right) \notag \\
&\quad + K \sin\left(2\pi \frac{t - T}{P} \right) 
\label{circular}
\end{align}

{is fitted to the data where the parameters are the semi-amplitude of the signal $K$, the orbital period $P$ and the time offset $T$. They were fitted the maximum a posterior (MAP) estimates of the parameters. Once no more significant signals are detected, we fit the final parameter values by taking the median value of their posterior distribution after computing an MCMC to the data similarly to how we did for the dET time series.

\par 

{As shown in Figure \ref{fig:gl480_spirou_RV_analysis}, the 9.5\,d signal reported in \citet{2020ApJS..250...29F} is also present in the SPIRou data. Before fitting this signal, the component from the Wapiti correction were added to the model, which reduced the RMS of the time series from 5.84 to 5.37\,m.s$^{-1}$ and removed a one-year signal from the data. Then we included this signal in the model, which led to a periodogram showing a long-term signal. This signal could be caused by a planet, the instrument, or a long-term magnetic cycle. To handle the long-term signal, we tested fitting it with a 1st- or 2nd-degree polynomial. The periodograms from these fits showed signals at 57\,d and 17.8\,d, respectively. We interpreted these signals as linked to stellar activity since they are close to the rotation period of the star  P$_{\mathbf{rot}}$ and its harmonic (P$_{\mathbf{rot}}$/3). Interestingly, when using a 2nd-degree polynomial, only the 17.8\,d signal appeared in the RV data, while no other harmonics are detected. However, this signal is also strong in the dLW and dET activity indicators (Figure \ref{fig:gl480_activity}). To further rule out the 17.8\,d signal as a possible planet, we computed a stacked periodogram centered at this period plus or minus 1 day and also studied how its log BF changed as the number of observations was increasing \citep{stackedperiodogram}. If the signal were from a planet, its significance would increase steadily. Instead, as seen in Figure \ref{fig:gl480_spirou_RV_analysis}, its behavior is erratic.}

\par 

{To correct for stellar activity in the RV data, we used a quasi-periodic GP. We also stopped including a polynomial term in the model, as the GP can correct for long-term trends while being statistically prefered (see Table \ref{BIC_models_480}). To model stellar activity accurately, we applied the priors from \citet{Camacho2023} for the GP's hyperparameters (Priors I). Here’s how we set the priors:}
\begin{itemize}
    \item {For the amplitude and jitter ($\sigma$) of the GP, we used a modified log-uniform distribution ($\mathcal{MLU}$; \citet{gregory2005}) with a 'knee' value of y$_\sigma$ (the standard deviation of the data) and an upper bound of 2y$_\mathbf{ptp}$ (twice the peak-to-peak difference of the RVs).}
    \item {For the smoothing parameter ($\Gamma$), we used a log-uniform distribution ($\mathcal{LU}$) with a range from 0.1 to 5.}
    \item{For the decaying timescale ($\lambda$), the prior ranged from $\delta$t$_\mathbf{av}$ (the average time between observations) to 10t$_\mathbf{span}$ (ten times the total time span of the observations).}
    \item {For the rotational period (P$_{\mathbf{rot}}$), we used a uniform distribution ranging from 1 to 1,000\,d.}
\end{itemize}

{The MAP estimate gave a P$_{\mathbf{rot}}$ value of 52\,d, consistent with the star’s known rotation period. Running an MCMC analysis gave a median  P$_{\mathbf{rot}}$ value of 26\,d (half the rotational period) with a smoothing parameter of 2 probably due to the complex dynamics behind the 17.8\,d signal and various local minima. To ensure  P$_{\mathbf{rot}}$ better represents the actual rotation period, we opted to use a different set of prior for the P$_{\mathbf{rot}}$, $\Gamma$ and $\lambda$ hyperparameters by using their posterior distribution from the MCMC analysis of the dET time series  (Priors II) or the dLW time series (Priors III). We chose these two indicators to create informative priors because the 17.8\,d signal is present in both of them.}

\par

{Once we obtained the hyper-parameters of the GP we keep searching for signals in the data by fixing its smoothing parameter $\Gamma$ and decay time $l$ while letting the amplitude coefficient free. The reason for this choice is to balance accuracy and computational efficiency. Simply subtracting the GP from the data before searching for a signal would not be appropriate because the GP's hyper-parameters would change when a signal with period P is added. Ideally, we should perform a MCMC analysis for each period in our grid, but this is too computationally intensive. Instead, we simplify the problem by keeping only the coefficient $A$ as a free variable which linearizes the problem, making the computation faster. If we do find a significant signal, we then recompute all the GP hyper-parameters after adding the signal and we verify that the new model remains statistically significant.}

{As shown in Figure \ref{fig:gl480_spirou_RV_analysis}, after correcting for activity, a 6.43\,d signal remains in the data for the three sets of priors. This period does not appear in the activity indicators and when modeled as a potential planetary signal, the MAP estimate of the semi-amplitude K is 1.50\,m.s$^{-1}$. However, since adding this signal also updated the hyperparameters of the GP, the log BF of this model compared to the previous one decreased to 2.31, 1.90 and 3.71 with respectively Priors I, II and III which is below our significance threshold. As a result, we cannot confidently claim the detection of a new 6.43\,d signal in Gl\,480 based on the SPIRou data alone. A summary of the models used and their BIC (MAP estimate) values can be seen in Table \ref{BIC_models_480}.}

\begin{figure*}
    \centering
    \includegraphics[width=0.75\linewidth]{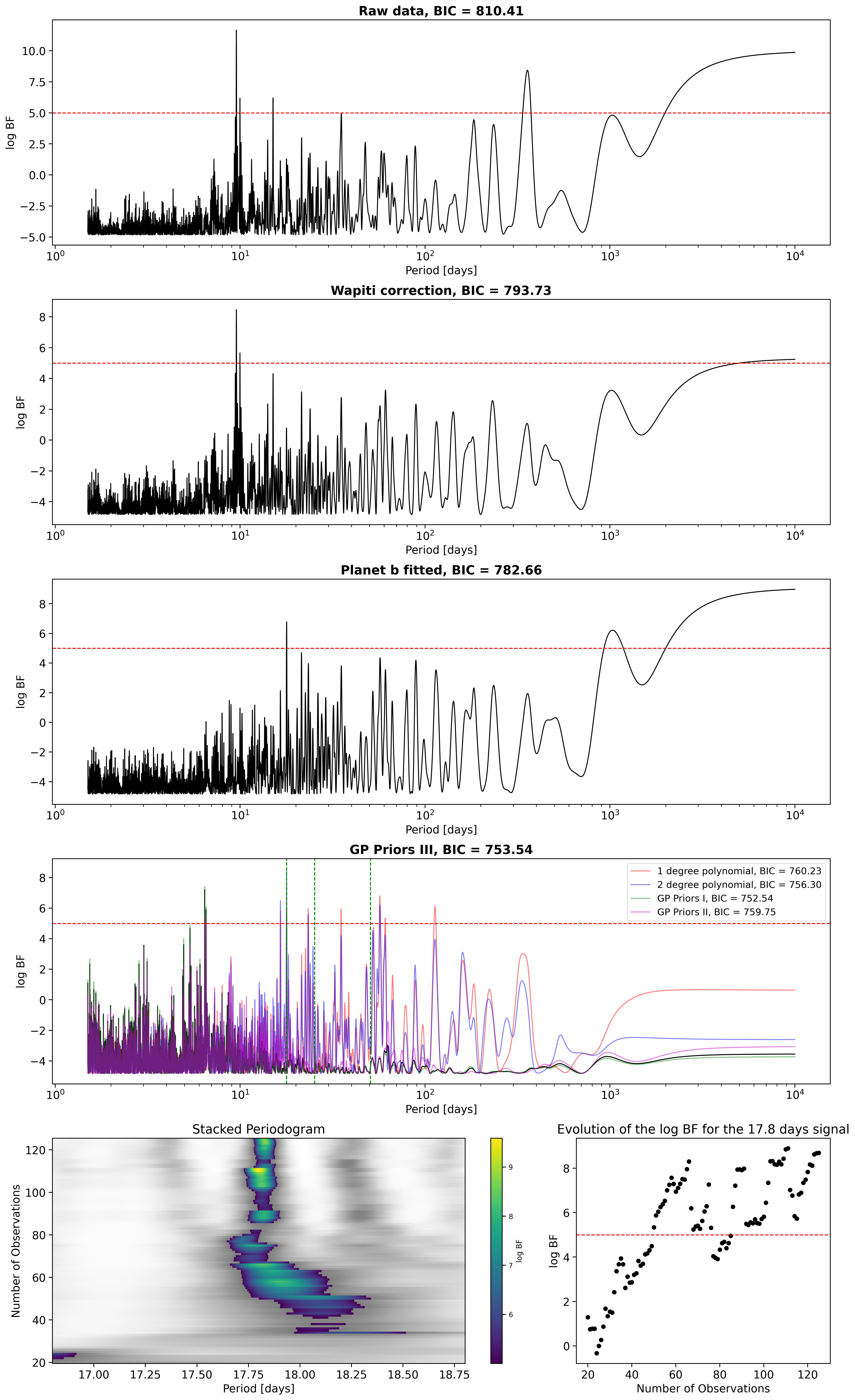} % Reduce size to fit columns
    \caption{Iterative search of signals with Bayesian periodograms using SPIRou data of Gl\,480. The top four panels display log BF periodograms for (a) raw data, (b) Wapiti correction, (c) Planet b model fit, and (d) long term and/or stellar activity correction. The dashed green lines in the bottom panel mark periodicities at the stellar rotation period (P$_{\mathbf{rot}}$), half rotation period (P$_{\mathbf{rot}}$/2), and one-third rotation period (P$_{\mathbf{rot}}$/3). The significance level of log BF $= 5$ is indicated by the horizontal red dashed line. The bottom left panel shows the stacked periodogram with log BF power values, where log BF values below 5 are in grey. The bottom right panel presents the variation of log BF as a function of the number of observations for the most significant period at 17.8\,d.}
    \label{fig:gl480_spirou_RV_analysis}
\end{figure*}

\subsubsection{Optical RV analysis}

{For the optical RV analysis, we used a combination of the HARPS and CARMENES dataset, which slightly differs from the analysis in \citet{2020ApJS..250...29F} that did not use the 7 CARMENES data points and because we discarded 20 HIRES RV measurements used in this study}. The same iterative search for signals  as for the SPIRou data was conducted. The known 9.5\,d signal is recovered and fitted leading to the detection of a significant signal at {53.64\,d} near the rotational period of the star which can be interpreted as of stellar activity origin (Figure \ref{fig:gl480_spirou_RV_analysis_optical}). Consequently we fitted a quasi-periodic GP to the data with a different instrumental jitter $\sigma$ for each instrument. We computed a MCMC on this model using the set of priors I.

\par 

{After removing stellar activity from the optical data, we found no other significant signals. The final MCMC analysis gives a semi-amplitude of  $5.0^{+0.5}_{-0.5}$\,m\,s$^{-1}$ for the 9.5\,d signal (see Table \ref{gl480_parameters}), which is noticeably different from the $6.80 \pm 0.87$\,m\,s$^{-1}$ value reported by \citet{2020ApJS..250...29F}. Without using a GP, we found a semi-amplitude of $5.5^{+0.7}_{-0.7}$\,m\,s$^{-1}$, suggesting that the difference in values is mostly due to excluding the HIRES data and adding a GP to our model. This semi-amplitude is also higher than the value found with SPIRou data alone of $3.6^{+0.5}_{-0.5}$\,m\,s$^{-1}$. However, this discrepancy is not caused by either the Wapiti correction or the GP applied on SPIRou data, as not using both still gives a semi-amplitude of $3.8^{+0.7}_{-0.7}$\,m\,s$^{-1}$. While this difference is within 5$\sigma$ and not necessarily incompatible, it is worth noting, as it could raise some uncertainty about the true planetary nature of this signal.}

\begin{figure}
    \centering
    \includegraphics[width=\linewidth]{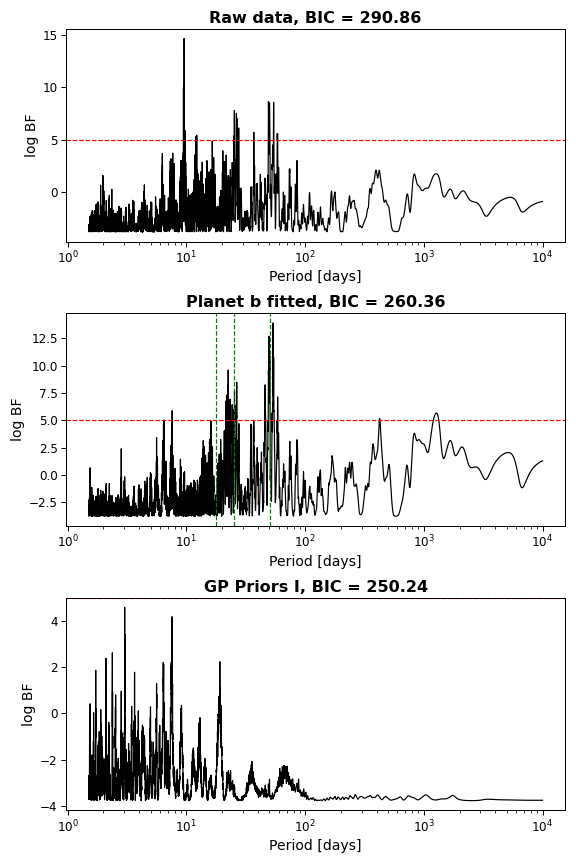} % Reduce size to fit columns
    \caption{Iterative search of signals with Bayesian periodograms using HARPS and CARMENES data of Gl\,480. The three panels display log BF periodograms for (a) raw data, (b) Planet b model fit, and (c) stellar activity correction. The dashed green lines in the bottom panel mark periodicities at the stellar rotation period (P$_{\mathbf{rot}}$), half rotation period (P$_{\mathbf{rot}}$/2), and one-third rotation period (P$_{\mathbf{rot}}$/3). The significance level of log BF $= 5$ is indicated by the horizontal red dashed line.}
    \label{fig:gl480_spirou_RV_analysis_optical}
\end{figure}

\subsubsection{Joint nIR-optical RV analysis}

{Finally, we combined the SPIRou data with the optical data to perform the final iterative search for signals, as shown in Figure \ref{fig:GL480_final_plot}. After correcting for the Wapiti systematics and the 9.5\,d signal, long-term signals appeared as four peaks at 958\,d, 1211\,d, 3006\,d, and 7322\,d.}

\par 

{However, as shown in the Figure \ref{fig:GL480_final_plot}, the individual datasets from different instruments do not overlap, and their time baselines are not above 1,000\,d. This makes it difficult to accurately estimate the properties of such a long-term signal, as they are heavily influenced by the choice of zero points for the RV datasets. Furthermore, this long-term signal was only seen in the SPIRou data (which has a long time span than both datasets) and not in the combined HARPS and CARMENES dataset, raising doubts about its planetary origin.}

\par 

{It is likely that this signal comes from SPIRou data alone, caused either by an instrumental effect or a long-term magnetic cycle affecting nIR RV measurements. This conclusion is supported by the fact that applying a GP to the SPIRou data alone is sufficient to correct for these long-term signals while providing the best fit when compared to circular models, as indicated by the BIC values shown in Table \ref{BIC_models_480}.}

\par

{We corrected for stellar activity using a GP with Priors I for the optical datasets and a different GP for the SPIRou data for which we tried the three sets of Priors. We found that Priors III were giving the best BIC value and we again detected a 6.4\,d signal in the residuals. When fitted, this signal had a log BF of 6.7, just above our detection threshold, indicating that adding the optical data slightly increased its significance and allowed for a detection. However, we note that with both Priors I and Priors II, the signal fell below the significance threshold. This suggests that additional observations are needed to confirm or rule out this low-amplitude signal.} 

\par

{For completeness, we included the 6.4\,d signal in the final model, but more data will be required to confidently confirm it as a new planetary signal orbiting Gl\,480. Our final model, therefore, includes the 9.5\,d planet, the 6.4d signal and GPs to model activity using Priors III for SPIRou data and Priors I for the optical data. We also tested whether an eccentric model for the signals provided a better fit by using a Keplerian model $\mu_K$ of the form}:

\begin{equation}
    \mu_K = \gamma + K \left[ \sin\left(\omega + \nu\left(t\right)\right) + e\cos\omega\right]
\end{equation}

{where the parameters of the model includes the semi-amplitude of the signal ($K$), the orbital period ($P$), the eccentricity $e$, the argument of periastron $\omega$, and the true anomaly $\nu$. The \texttt{radvel} package \citep{2018PASP..130d4504F} was used to model the Keplerian and we found that considering an eccentric model did not improve the fitting of our data. Therefore, we chose a circular model for both signals, as it had the lowest BIC.}

\begin{table*}
\centering
\caption[]{Comparison of models applied on Gl\,480 with different datasets using their BIC values.}
\small
\begin{tabular}{c c c}
\hline
\noalign{\smallskip}
Dataset & Model & BIC (MAP estimate) \\
\noalign{\smallskip}
\hline
\noalign{\smallskip}
SPIRou & offset & 810\\
 & offset + Wapiti & 794\\
 & offset + Wapiti + 9.5d & 783 \\
 & offset + Wapiti + 9.5d + 1 degree polynomial & 760 \\
 & offset + Wapiti + 9.5d + 2 degree polynomial & 756 \\
 & offset + Wapiti + 9.5d + GP Priors I & 753 \\
 & offset + Wapiti + 9.5d + GP Priors II & 760 \\
 & \textbf{offset + Wapiti + 9.5d + GP Priors III} & \textbf{754} \\
 & offset + Wapiti + 9.5d + GP Priors I + 6.4d & 748 \\
 & offset + Wapiti + 9.5d + GP Priors II + 6.4d & 756 \\
 & offset + Wapiti + 9.5d + GP Priors III + 6.4d & 746 \\
\noalign{\smallskip}
\hline
\noalign{\smallskip}
HARPS + CARMENES & offsets & 291 \\
 & offsets + 9.5d & 260 \\
 & \textbf{offsets + 9.5d + GP Priors I} & \textbf{250} \\
\noalign{\smallskip}
\hline
\noalign{\smallskip}
ALL & offset & 1107 \\
 & offset + Wapiti & 1091 \\
 & offset + Wapiti + 9.5d & 1050 \\
 & offset + Wapiti + 9.5d + 957d & 1039 \\
 & offset + Wapiti + 9.5d + 1211d & 1027 \\
 & offset + Wapiti + 9.5d + 3006d & 1036 \\
 & offset + Wapiti + 9.5d + 7322d & 1036 \\
 & offset + Wapiti + 9.5d + SPIRou GP Priors I + Optical GP Priors I & 1016 \\
 & offset + Wapiti + 9.5d + SPIRou GP Priors II + Optical GP Priors I & 1017 \\
 & offset + Wapiti + 9.5d + SPIRou GP Priors III + Optical GP Priors I & 1010 \\
 & offset + Wapiti + 9.5d ecc. + SPIRou GP Priors III + Optical GP Priors I & 1069 \\
 & \textbf{offset + Wapiti + 9.5d + SPIRou GP Priors III + Optical GP Priors I + 6.4d} & \textbf{997} \\
 &offset + Wapiti + 9.5d + SPIRou GP Priors III + Optical GP Priors I + 6.4d ecc. & 1025 \\
\noalign{\smallskip}
\hline
\end{tabular}
\begin{minipage}{\textwidth} % This ensures the note is centered and spans the table width
\small
The bold text indicates the final model.
\end{minipage}\label{BIC_models_480}
\end{table*}

\par 

{Our combined nIR and optical radial velocity RV analysis of Gl\,480 confirms the presence of the 9.5\,d signal. Although the nIR and optical datasets individually yield slightly different semi-amplitude values, the final combined analysis results in a semi-amplitude of $K = 4.5^{+0.3}_{-0.3}$\,m\,s$^{-1}$. A detailed comparison of the semi-amplitude and period posterior distributions across the datasets is provided in Appendix \ref{sec:RVanalysis480}. Additionally, we detect a low-amplitude signal at 6.4\,d in the combined dataset. This signal barely exceeds our detection threshold and it is only the case when a specific set of priors is applied, rendering it quite weak. Nonetheless, we include this signal in the final model for consistency with our detection criterion, with an associated semi-amplitude of $K = 1.6^{+0.3}_{-0.3}$\,m\,s$^{-1}$.}

\par 

{However, although this signal was marginally detected in the SPIRou data alone, it was entirely absent from the optical datasets. To investigate its non-detection in the HARPS and CARMENES datasets, we performed a simple injection-recovery test similar to the one conducted in \citet{2023A&A...670A...5S}. First, we removed the 9.5\,d planetary signal from the data and injected 10,000 sinusoidal signals with a semi-amplitude K$_\mathbf{sim}$ $\sim$ $\mathcal{N}\left(1.6, 0.3\right)$, a period P$_\mathbf{sim}$ $\sim$ $\mathcal{N}\left(6.4349, 0.0006\right)$  (corresponding to the values obtained from the MCMC analysis in Table \ref{gl480_parameters}), and a random phase. We then subtracted the stellar activity using the same GP hyperparameters as those measured for the one-planet model, while recomputing the activity model for each individual dataset.}

\par 

{Next, we computed the Bayesian periodogram at the injected period and assessed whether it could be recovered. The results, shown in Figure \ref{fig:GL480_log_BF_optical}, display the histogram of the log BF values for the injected period. Only 19\% of the simulations successfully detected the signal with a log BF above 5, and the mean value is of 2.0 close to the value of 1.8 retrieved from the residuals of the actual datasets. Thus, our injection-recovery test indicates that the non-detection of the 6.4\,d signal in the optical datasets seems compatible with the SPIRou candidate detection .}

\begin{figure}
    \centering
    \includegraphics[width=\linewidth]{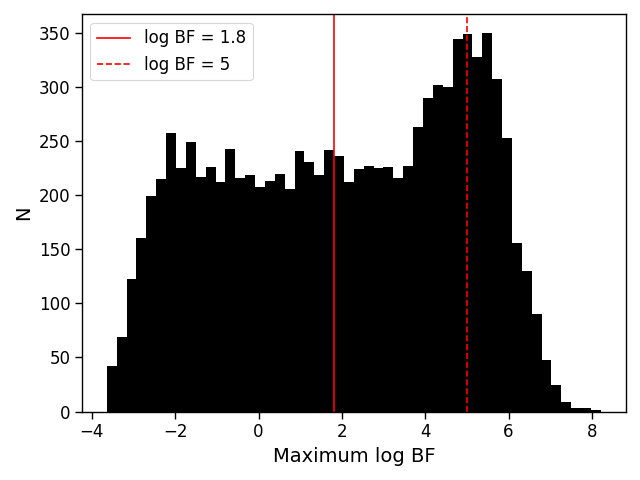} % Reduce size to fit columns
    \caption{Histogram of the maximum log BF values of the injected period in the injection-recovery simulations. The solid red line indicates the log BF value of 1.8 retrieved from the residuals of the real datasets, while the dashed red line marks the detection threshold at log BF = 5.}
    \label{fig:GL480_log_BF_optical}
\end{figure}

\par 

Finally, we estimated the minimum planet's mass, $\mathrm{M_p} \sin i$, by using the star's mass and its uncertainty obtained from \cite{2022MNRAS.516.3802C}, which is listed in Table \ref{stellar_properties}. To estimate the equilibrium temperature ($\mathrm{T}_\mathrm{eq}$), we used the star's radius and effective temperature also derived from the same study. The corner plot from the MCMC computed over the data sets are available in Appendix \ref{sec:RVanalysis480}. The final fit on the dataset using all instruments is presented in Figure \ref{fig:GL480_final_plot}, we provide the phase-folded representation of the data and the residuals.

\begin{figure*}
    \centering
    \includegraphics[width=\linewidth]{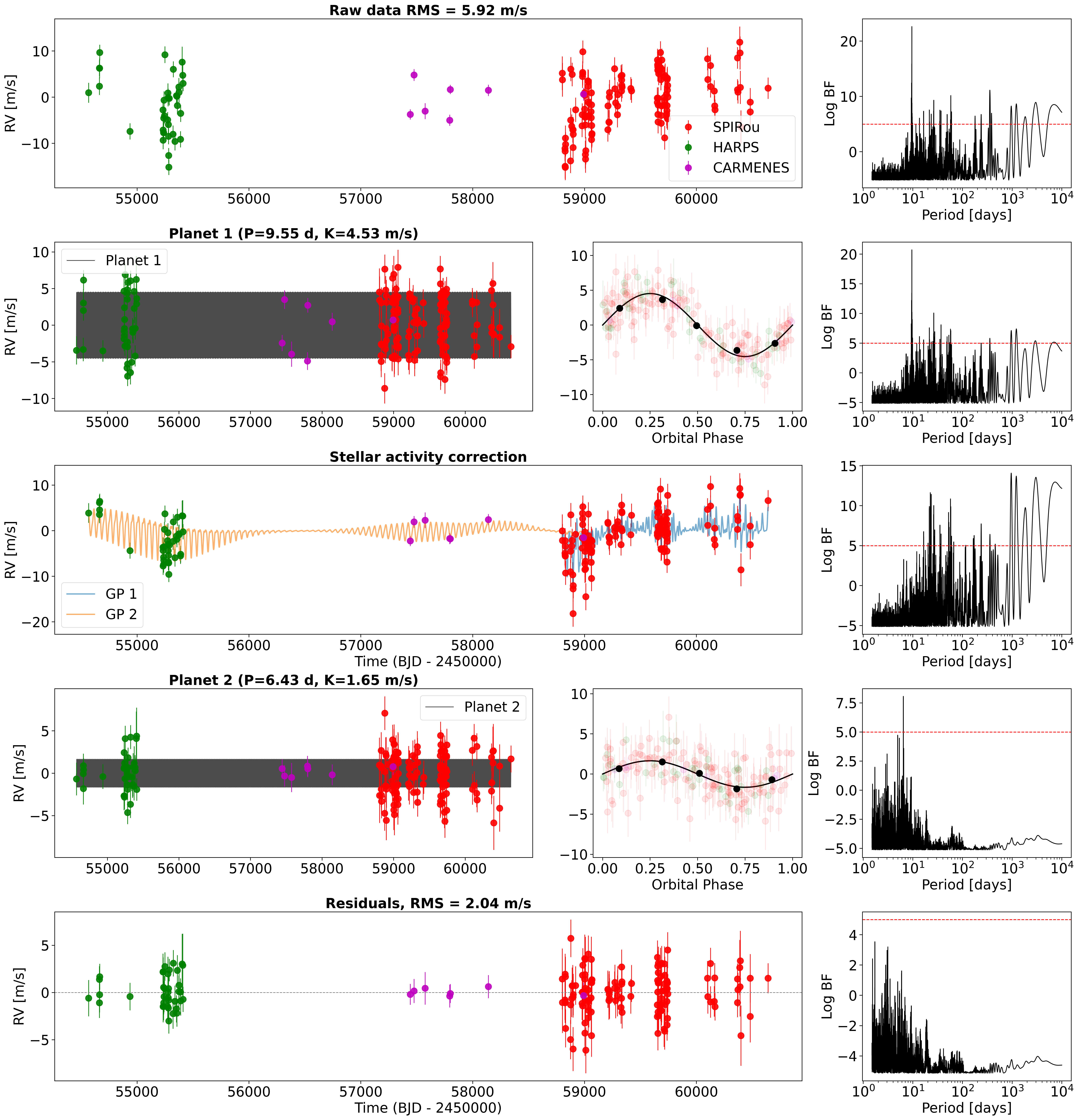} % Reduce size to fit columns
    \caption{Final RV model for Gl\,480. The top panel shows the raw RV data with offsets removed, plotted for SPIRou, HARPS, and CARMENES instruments (RMS = 5.92\,m.s$^{-1}$). The second and fourth rows display the contributions of planets 1 and 2 to the RV signal, respectively, along with their phase-folded RV curves. The third row illustrates the correction for stellar activity using GPs, with individual GP contributions shown for each grouped instrument. The bottom panel presents the residuals after removing all modeled signals (RMS = 2.04\,m.s$^{-1}$). The rightmost column contains log BF periodograms highlighting significant periods in the RV data, the significance level of 5 is indicated by the horizontal red dashed line.}
    \label{fig:GL480_final_plot}
\end{figure*}

\begin{table*}
\centering
\caption{Orbital and derived planetary parameters from the analysis of Gl\,480.}
\small % Reduce font size
\begin{tabular}{cccccc}
\hline
& SPIRou & Optical & ALL & Priors & Feng et al. (2020)  \\ [0.5ex]
\hline
\noalign{\smallskip}
\multicolumn{6}{c}{\textbf{Instrumental Parameters}} \\
\noalign{\smallskip}
\hline
\noalign{\smallskip}
$\gamma_{SPIRou}$ (m\,s$^{-1}$) & $-4176.4^{+1.1}_{-1.0}$ & - &  $-4176.1^{+1.0}_{-1.0}$ & $\mathcal{U}\left(-10^{10}, 10^{10}\right)$ & - \\ 
$\gamma_{CARMENES}$ (m\,s$^{-1}$) & - & $0.3^{+3.0}_{-3.3}$ & $0.05^{+2.5}_{-2.8}$ & $\mathcal{U}\left(-10^{10}, 10^{10}\right)$   & - \\ 
$\gamma_{HARPS}$ (m\,s$^{-1}$) & - & $2.1^{+3.4}_{-3.0}$ & $2.1^{+2.8}_{-2.8}$ & $\mathcal{U}\left(-10^{10}, 10^{10}\right)$   & - \\ 
$\sigma_{SPIRou}$ (m\,s$^{-1}$) & $2.4^{+0.4}_{-0.5}$ & - & $1.9^{+0.5}_{-0.4}$ & $\mathcal{MLU}\left(y_\sigma, 2y_{ptp}\right)$ &  - \\ 
$\sigma_{CARMENES}$ (m\,s$^{-1}$) & - &$0.1^{+0.5}_{-0.1}$ & $0.08^{+0.5}_{-0.1}$ & $\mathcal{MLU}\left(y_\sigma, 2y_{ptp}\right)$ &  - \\ 
$\sigma_{HARPS}$ (m\,s$^{-1}$) & - & $1.6^{+0.6}_{-0.6}$ & $0.8^{+0.7}_{-0.7}$ & $\mathcal{MLU}\left(y_\sigma, 2y_{ptp}\right)$ & - \\ 
\noalign{\smallskip}
\hline
\noalign{\smallskip}
\multicolumn{6}{c}{\textbf{Wapiti Systematic}} \\
\noalign{\smallskip}
\hline
\noalign{\smallskip}
$a_1$ (m/s) & $16.6^{+8.0}_{-8.1}$ & - & $18.0^{+7.5}_{-7.8}$ & $\mathcal{U}\left(-10^{2},10^2\right)$ \\ 
\noalign{\smallskip}
\hline
\noalign{\smallskip}
\multicolumn{6}{c}{\textbf{GP Parameters}} \\
\noalign{\smallskip}
\hline
\noalign{\smallskip}
$A_{SPIRou}$ (m\,s$^{-1}$) & $4.3^{+2.9}_{-1.3}$ & - & $4.4^{+0.8}_{-0.6}$ & $\mathcal{MLU}\left(y_\sigma, 2y_{ptp}\right)$ & -\\
$\lambda_{SPIRou}$ (days) & $563.6^{+1049.2}_{-359.8}$ & - & $69.4^{+20.8}_{-19.8}$ & Posterior$_{\mathrm{dLW}}$ & -  \\
$\Gamma_{SPIRou}$  & $1.0^{+1.4}_{-0.7}$ & - & $4.9^{+3.6}_{-2.3}$ & Posterior$_{\mathrm{dLW}}$ & - \\
$P_{rot, SPIRou}$ (days) & $49.7^{+0.8}_{-0.7}$ & - & $51.6^{+1.2}_{-1.4}$ & Posterior$_{\mathrm{dLW}}$ & - \\
$A_{Optical}$ (m\,s$^{-1}$) & - & $4.3^{+2.9}_{-1.3}$ & $4.2^{+2.6}_{-1.2}$ & $\mathcal{MLU}\left(y_\sigma, 2y_{ptp}\right)$ & -\\
$\lambda_{Optical}$ (days) & - & $563.6^{+1049.2}_{-359.8}$ & $438^{+613}_{-269}$ & $\mathcal{LU}\left(\delta t_{av}, 10t_{span}\right)$ & -  \\
$\Gamma_{Optical}$  & - & $1.0^{+1.4}_{-0.7}$ & $1.3^{+2.0}_{-0.9}$ & $\mathcal{LU}\left(.1, 5\right)$ & - \\
$P_{rot, Optical}$ (days) & - & $49.7^{+0.8}_{-0.7}$ & $50.0^{+0.8}_{-0.8}$ & $\mathcal{U}\left(1, 10^{3}\right)$   & - \\ 
\noalign{\smallskip}
\hline
\noalign{\smallskip}
\multicolumn{6}{c}{\textbf{Orbital Parameters}} \\
\noalign{\smallskip}
\hline
\noalign{\smallskip}
$Pb$ (days) & $9.548^{+0.005}_{-0.005}$ & $9.553^{+0.003}_{-0.003}$ & $9.5536^{+0.0006}_{-0.0006}$ & $\mathcal{U}\left(1,10^{4}\right)$ & $9.567 \pm 0.005$ \\
$Kb$ (m\,s$^{-1}$) & $3.6^{+0.5}_{-0.5}$ & $5.0^{+0.5}_{-0.5}$ & $4.5^{+0.3}_{-0.3}$ & $\mathcal{U}\left(0,2y_{ptp}\right)$ & $6.80 \pm 0.87 $ \\
$Tb$ (MJD) & $59430.1^{+0.2}_{-0.2}$ & $55627.8^{+0.2}_{-0.2}$ & $58446.0^{+0.1}_{-0.1}$ & $\mathcal{U}\left(-10^{10}, 10^{10}\right)$ & - \\
$e_b$  & - & - & - & - & $0.10 \pm 0.7$ \\
$Pc$ (days) & - & - & $6.4349^{+0.0006}_{-0.0006}$ & $\mathcal{U}\left(1,10^4\right)$ & - \\
$Kc$ (m\,s$^{-1}$) & - & - & $1.6^{+0.3}_{-0.3}$ & $\mathcal{U}\left(0,2y_{ptp}\right)$ & - \\
$Tc$ (MJD) & - & - & $58443.3^{+0.2}_{-0.2}$ & $\mathcal{U}\left(-10^{10}, 10^{10}\right)$ & - \\
\noalign{\smallskip}
\hline
\noalign{\smallskip}
\multicolumn{6}{c}{\textbf{Derived Parameters}} \\
\noalign{\smallskip}
\hline
\noalign{\smallskip}
$\mathrm{M_{p, b}} \sin i$ (M$_{\oplus}$) & $7.1^{+1.0}_{-1.0}$ & $9.7^{+1.1}_{-1.0}$ & $8.8^{+0.7}_{-0.7}$ & - & $13.2 \pm 1.7$ \\ 
$a_b$ (AU) & $0.068^{+0.001}_{-0.001}$ & $0.068^{+0.001}_{-0.001}$ & $0.068^{+0.001}_{-0.001}$ & - & $0.068 \pm 0.001$ \\ 
T$_\mathrm{eq, b}$ (K) & $406^{+6}_{-6}$ & $406^{+6}_{-6}$ & $406^{+6}_{-6}$  & - & - \\ 
$\mathrm{M_{p, c}} \sin i$ (M$_{\oplus}$) & - & - & $2.7^{+0.6}_{-0.6}$ & - & - \\ 
$a_c$ (AU) & - & - & $0.052^{+0.001}_{-0.001}$ & - & - \\ 
T$_\mathrm{eq, c}$ (K) & - & - & $463^{+7}_{-7}$  & - & - \\ 
\noalign{\smallskip}
\hline
\noalign{\smallskip}
\end{tabular}
\begin{minipage}{\textwidth} % This ensures the note is centered and spans the table width
\small
The parameters derived from the \texttt{Wapiti} corrected RV data from SPIRou are compared to those from \citet{2020ApJS..250...29F} using HARPS and HIRES spectrographs. The parameters derived by combining all data sets including the few CARMENES data correspond to the ALL column. We also show for completeness the eccentricity $e$ and argument of periastron $\omega$ from the Keplerian fit conducted in \citet{2020ApJS..250...29F}.
\end{minipage}
\label{gl480_parameters}
\end{table*}

\subsection{Gl\,382}

\subsubsection{Stellar activity}

For Gl\,382, we employed the same analysis approach used for Gl\,480. Initially, we examined various indicators of stellar activity available for both HARPS and SPIRou, just like we did for Gl\,480. Additionally, we had access to a larger dataset from CARMENES, comprising 77 data points, which allowed us to use multiple stellar activity indicators. These indicators included the dLW, the H$\alpha$ and NaD1 indices, as well as the Ca II IRT1 (849.802,nm), Ca II IRT2 (854.209,nm), and Ca II IRT3 (866.214,nm) indices.

\par 

The periodograms {of those stellar activity indicators} are presented in Figure \ref{fig:gl382_activity}. In these figures, we have highlighted the rotational period in green, its first harmonic.

\par 

Upon an examination of these indicators, we found that both CARMENES and HARPS data exhibited power in their H$\alpha$ periodograms near the known 21.3 - 21.9\,day rotational period. {Moreover, for the CARMENES data, the Ca II IRT 1, 2, and 3 indices also show a significant period corresponding to the rotational period. In HARPS data the rotational period is also observed in the NaD1 at a more significant level than in CARMENES data, a 50\,d signal is also observed in HARPS dLW and  H$\alpha$ activity proxies.}

\par

As for the SPIRou indicators, {we again used the dLW, B$_\ell$ and dET time series. The dLW time series is contaminated by a long term signal but the B$_\ell$ and dET time series allow to detect the known rotational period. However we observed that in the latter time series, as for HARPS data, a 50\,d signal is present at a significance similar to the rotational period, this signal is also highlighted by green dashed line in the Figure \ref{fig:gl382_activity}. The origin of this signal is unclear however it is unlikely to be due to an alias of the rotatinoal period or its harmonics since it would otherwise also be observed in the B$_\ell$ timeseries, all the more so as it is also observed in HARPS indicators. To characterize the rotational period we computed a MCMC over the dET time series using quasi-periodic GP which gave us an estimated rotational period for Gl\,382 of {$\mathrm{P_{rot}} = 21.41^{+0.14}_{-0.14}$\,d}. The posterior distribution of the hyper-parameters can be seen in the corner plot in Appendix \ref{sec:RVanalysis382}.}

\begin{figure*}[!htbp]
    \centering
    \includegraphics[width=\linewidth]{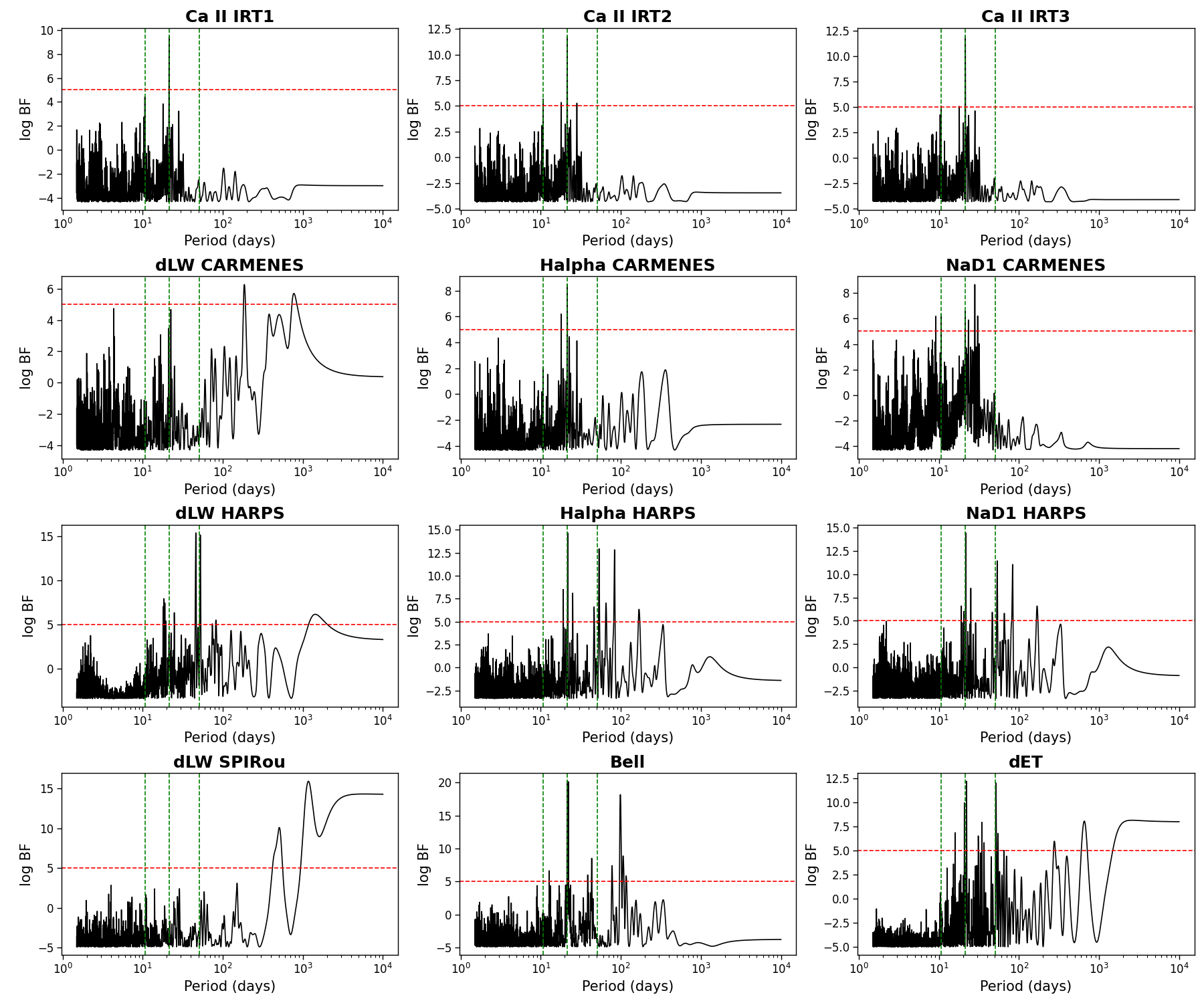}
    \caption{Bayesian periodograms of various stellar activity indicators of Gl\,382. In order the Ca II IRT 1, 2, 3, dLW, $H\alpha$ and the NaD1 index time series from CARMENES data; the dLW, $H\alpha$ and the NaD1 index time series from HARPS data; and the dLW, B$\ell$ and dET time series from SPIRou data. The dashed green lines in the bottom panel mark periodicities at the stellar rotation period (P$_{\mathbf{rot}}$), half rotation period (P$_{\mathbf{rot}}$/2), and at 51.1\,d.  The significance level of log BF $= 5$ is indicated by the horizontal red dashed line.}
    \label{fig:gl382_activity}
\end{figure*}

\subsubsection{SPIRou RV analysis}

{Starting with the analysis of our SPIRou data, Figure \ref{fig:GL382_bfs_analysis_spirou} shows significant long-term signals that are effectively removed after applying the Wapiti correction. This correction isolates a 10.4\,d signal, which we attribute to stellar activity due to its closeness to the first harmonic of the star's rotational period. Interestingly, no signal corresponding to the actual rotational period is detected, making Gl\,382 another example of a star exhibiting a stable rotational harmonic \citep{Ahrer2021}.}

\par 

{We corrected for stellar activity following the approach described in the previous section, employing a GP with the set of Priors I. After modeling the activity, we observed that the P$_{rot}$ hyperparameter converged to a value of 20.8\,d, which is slightly different from the known value of 21.14\,d. We also detected a 51.1\,d signal in our data (Figure \ref{fig:GL382_bfs_analysis_spirou}). However, this signal is also present in ancillary activity indicators, including SPIRou's dET time series as well as in the H$\alpha$ time series from HARPS. Additionally, a stacked periodogram centered around this period (Figure \ref{fig:GL382_bfs_analysis_spirou}) shows that the signal's significance does not gradually increase with additional observations but instead emerges abruptly due to the most recent data points. Given this behavior, we attributed it to stellar activity, even though its physical origin is unclear as mentioned earlier. To further correct for this activity, we applied an alternative prior set, Priors II, which incorporate the posterior distributions from the MCMC analysis of the dET time series as priors for the smoothing parameter $\Gamma$, and the decay timescale $\lambda$ which can be seen in Appendix \ref{sec:RVanalysis382}. For the rotational period, we left a uniform prior since we know the value should be close to 20.8\,d, whereas the posterior distribution of the dET time series would incorrectly center it around 21.4\,d. This set of priors successfully produced a GP model that effectively removed both the 10.4\,d and 51\,d signals from the data. Therefore, we adopted Priors II for the SPIRou dataset, and after applying this GP-based activity correction, no significant signals remained in the residuals.}

\begin{figure*}[!htbp]
    \centering
    \includegraphics[width=0.85\linewidth]{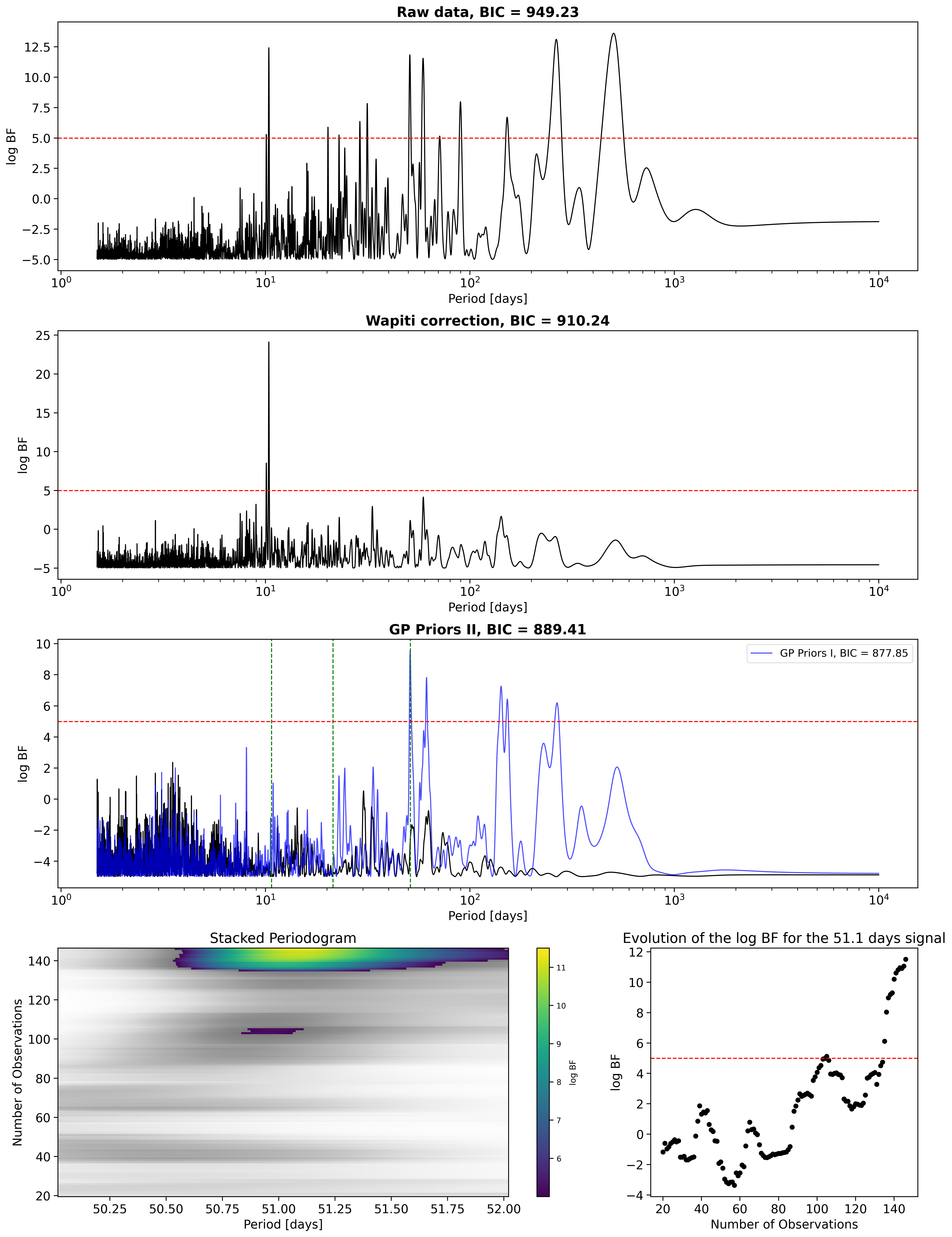} % Reduce size to fit columns
    \caption{Iterative search of signals with Bayesian periodograms using SPIRou data of Gl\,382. The top four panels display log BF periodograms for (a) raw data, (b) Wapiti correction, and (c) stellar activity correction. The dashed green lines in the bottom panel mark periodicities at the stellar rotation period (P$_{\mathbf{rot}}$), half rotation period (P$_{\mathbf{rot}}$/2), and 51\,d period. The significance level of log BF $= 5$ is indicated by the horizontal red dashed line. The bottom left panel shows the stacked periodogram with log BF power values, where log BF values below 5 are in grey. The bottom right panel presents the variation of log BF as a function of the number of observations for the most significant period at 51\,d.}
    \label{fig:GL382_bfs_analysis_spirou}
\end{figure*}

\subsubsection{Optical and joint nIR-optical RV analysis}

{In the analysis of the CARMENES and HARPS RV time series (Figure \ref{fig:gl382_spirou_RV_analysis_optical}), we detected a prominent 10.4\,d signal, which corresponds to the first harmonic of the star's rotational period. Notably, in the optical data, we also observed a signal at 21.5\,d. To correct for stellar activity, we applied a GP model using the set of Priors I. After this correction, a weak 43.4\,d signal emerged in the residuals, with a log BF of 5.36. However, as mentioned earlier, this periodogram was generated for computational efficiency by allowing only the GP hyperparameter $A$ to vary while keeping the others fixed. When we include this signal in our model and update all parameters, including the full set of GP hyperparameters, the log BF decreases to 1.06, leading us to dismiss this signal. Interestingly, the 51\,d signal observed in the SPIRou data is entirely absent from the RV residuals of this optical dataset, corroborating our claim that this signal was originating from stellar activity. Finally, we incorporated the SPIRou dataset to conduct a joint nIR and optical RV analysis. The SPIRou data were corrected for stellar activity using the set of Priors II. Despite combining all datasets, no significant signals remained in the residuals after correction. The BIC values of the different models used can be seen in Table \ref{BIC_models_382}.}

\begin{figure}
    \centering
    \includegraphics[width=\linewidth]{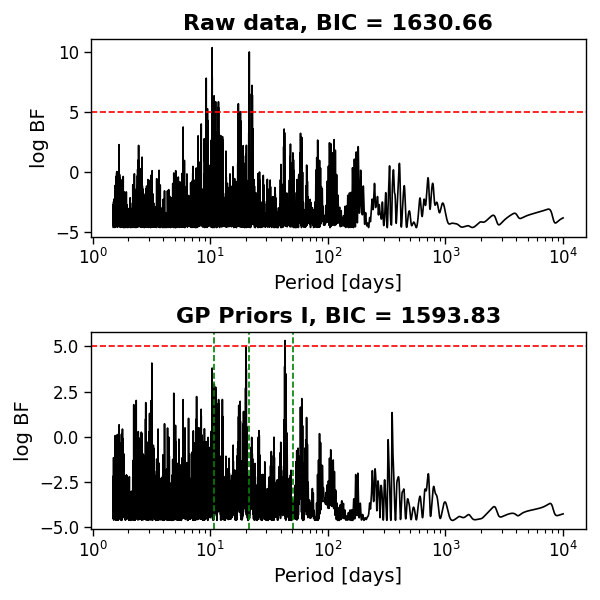} % Reduce size to fit columns
    \caption{Iterative search of signals with Bayesian periodograms using HARPS and CARMENES data of Gl\,382. The twp panels display log BF periodograms for (a) raw data, (b) stellar activity correction. The dashed green lines in the bottom panel mark periodicities at the stellar rotation period (P$_{\mathbf{rot}}$), half rotation period (P$_{\mathbf{rot}}$/2), and 51 day period. The significance level of log BF $= 5$ is indicated by the horizontal red dashed line.}
    \label{fig:gl382_spirou_RV_analysis_optical}
\end{figure}

\begin{table*}
\centering
\caption[]{Comparison of models applied on Gl\,382 with different datasets using their BIC values.}
\small
\begin{tabular}{ccc}
\hline
\noalign{\smallskip}
Dataset & Model & BIC (MAP estimate) \\
\noalign{\smallskip}
\hline
\noalign{\smallskip}
SPIRou & offset & 949\\
 & offset + Wapiti & 910\\
 & offset + Wapiti + GP Priors I & 878 \\
 & \textbf{offset + Wapiti + GP Priors II} & \textbf{889} \\
\noalign{\smallskip}
\hline
\noalign{\smallskip}
HARPS + CARMENES & offsets & 1631 \\
 & \textbf{offsets + GP Priors I} & \textbf{1594} \\
\noalign{\smallskip}
\hline
\noalign{\smallskip}
ALL & offset & 1631 \\
 & offset + Wapiti & 1594 \\
 & \textbf{offset + Wapiti + SPIRou GP Priors II + Optical GP Priors I} & \textbf{1547} \\
\noalign{\smallskip}
\hline
\end{tabular}
\begin{minipage}{\textwidth} % This ensures the note is centered and spans the table width
\small
The bold text indicates the final model.
\end{minipage}\label{BIC_models_382}
\end{table*}

{The absence of a detected signal after correcting for stellar activity still allows us to estimate detection limits for potential additional companions that may have remained undetected in this study. To do so, we performed another injection-recovery test, similar to the one conducted previously for Gl\,480, with the key difference that this time we used a semi-amplitude grid linearly spaced from 0.1 to 3\,m.s$^{-1}$ and an orbital period grid linearly spaced from 40 to 150 days, both with 100 steps. For each combination of semi-amplitude and period, we injected 10 signals with a phase linearly sampled from 0 to 2$\pi$.}

\par 

{The chosen period range is because we estimated that the habitable zone (HZ) of this star extends from 0.19 to 0.41 AU \citep{hz}, corresponding to orbital periods of approximately 45.7 days and 134.3 days, respectively. We then assessed whether the injected signals could be recovered, with the results shown in Figure \ref{fig:GL382_log_BF_injection_recovery}.}

\par 

{To estimate the upper limit on the detectable semi-amplitude of a potential companion in the habitable zone, we identified the semi-amplitude values for which the detection rate reached 100\%, from which we calculated the corresponding minimum mass. The histograms, presented in Figure \ref{fig:GL382_log_BF_injection_recovery}, indicate an upper limit on the semi-amplitude of a potential companion of K$_\mathbf{upper} = 1.3 \pm 0.3$\,m.s$^{-1}$. This suggests that any potential companion within the habitable zone of Gl\,382 would have a minimum mass below M$_\mathbf{upper} = 5.9 \pm 0.8$\,M$_\oplus$.}

\begin{figure*}
    \centering
    \includegraphics[width=0.7\linewidth]{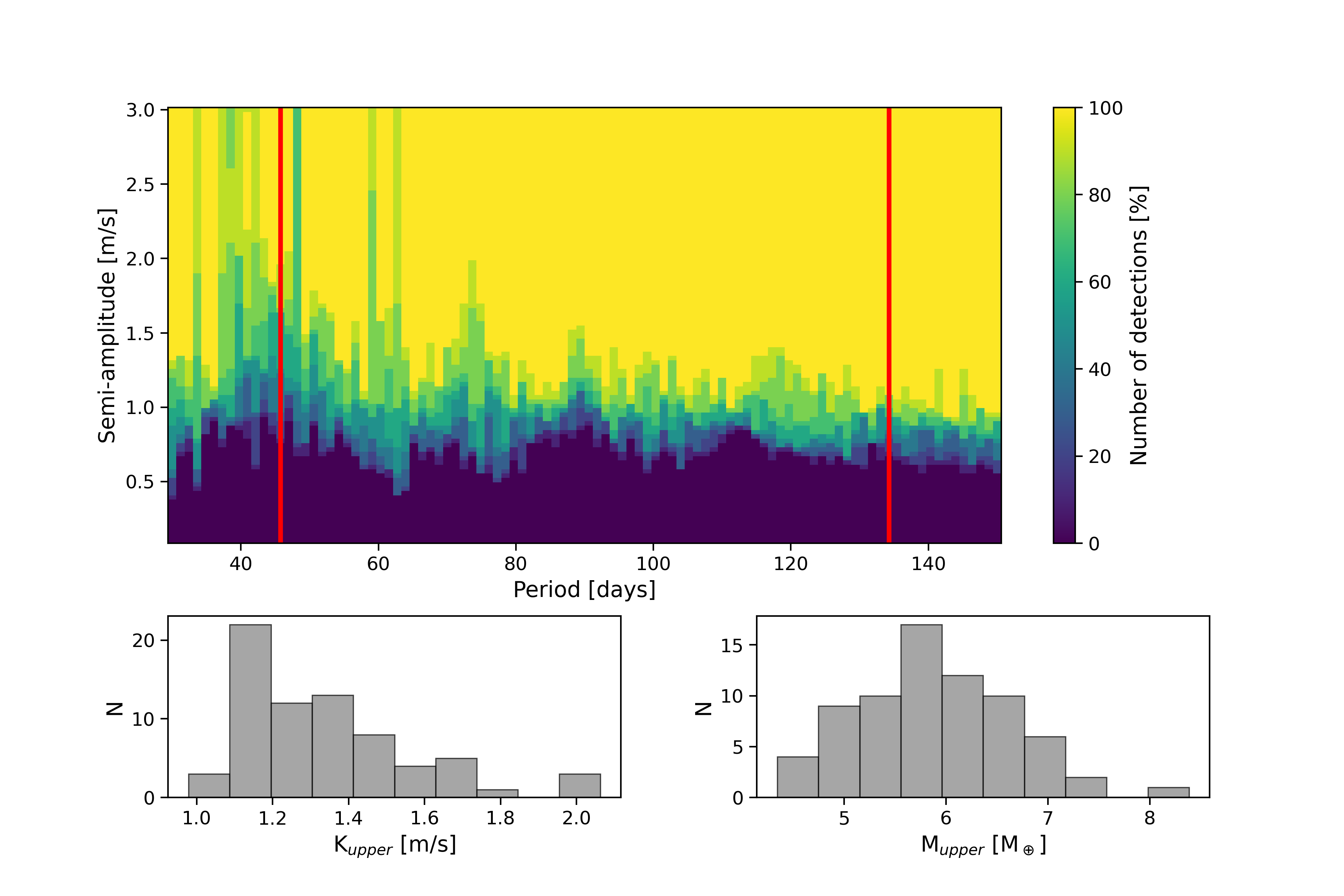} % Reduce size to fit columns
    \caption{Results of the injection-recovery test assessing the detectability of a planetary signal in the dataset. The top panel shows the detection rate as a function of orbital period and semi-amplitude, with colors representing the percentage of successful detections. The red vertical lines indicate the boundaries of the habitable zone. The bottom panels display histograms of K$_\mathbf{upper}$ (left) and M$_\mathbf{upper}$ (right), representing the semi-amplitude threshold at which the detection rate reaches 100\% and its corresponding minimum planetary mass.}
    \label{fig:GL382_log_BF_injection_recovery}
\end{figure*}

{In summary, stellar activity significantly impacts the RV data of Gl\,382, but its effect on the data differs between the optical and nIR domains. In the optical, stellar activity creates signals at both the rotational period and its first harmonic, with the latter being more pronounced. Conversely, in the nIR, only the first harmonic is strongly present. Furthermore, if stellar activity is not well corrected, an additional signal near 50\,d, likely of stellar origin, can appear in the residuals. However, this does not necessarily imply that this effect occurs exclusively in the nIR, as the optical and nIR datasets are not simultaneous. Activity could still influence both in a similar manner, especially given that the GP amplitudes are comparable. After properly correcting for stellar activity across all RV datasets, no significant signals are present and the final model can be seen in Figure \ref{fig:gl382_spirou_RV_analysis} as well as the parameters values in Table \ref{gl382_parameters}. However, the different ways in which stellar activity manifests in Gl\,382 across different wavelengths is intriguing. Although exploring the origin of the 50\,d signal is beyond the scope of this study, Gl\,382 could be an interesting playground for developing and testing new methods for stellar activity correction.}

\begin{figure*}[!htbp]
    \centering
    \includegraphics[width=\linewidth]{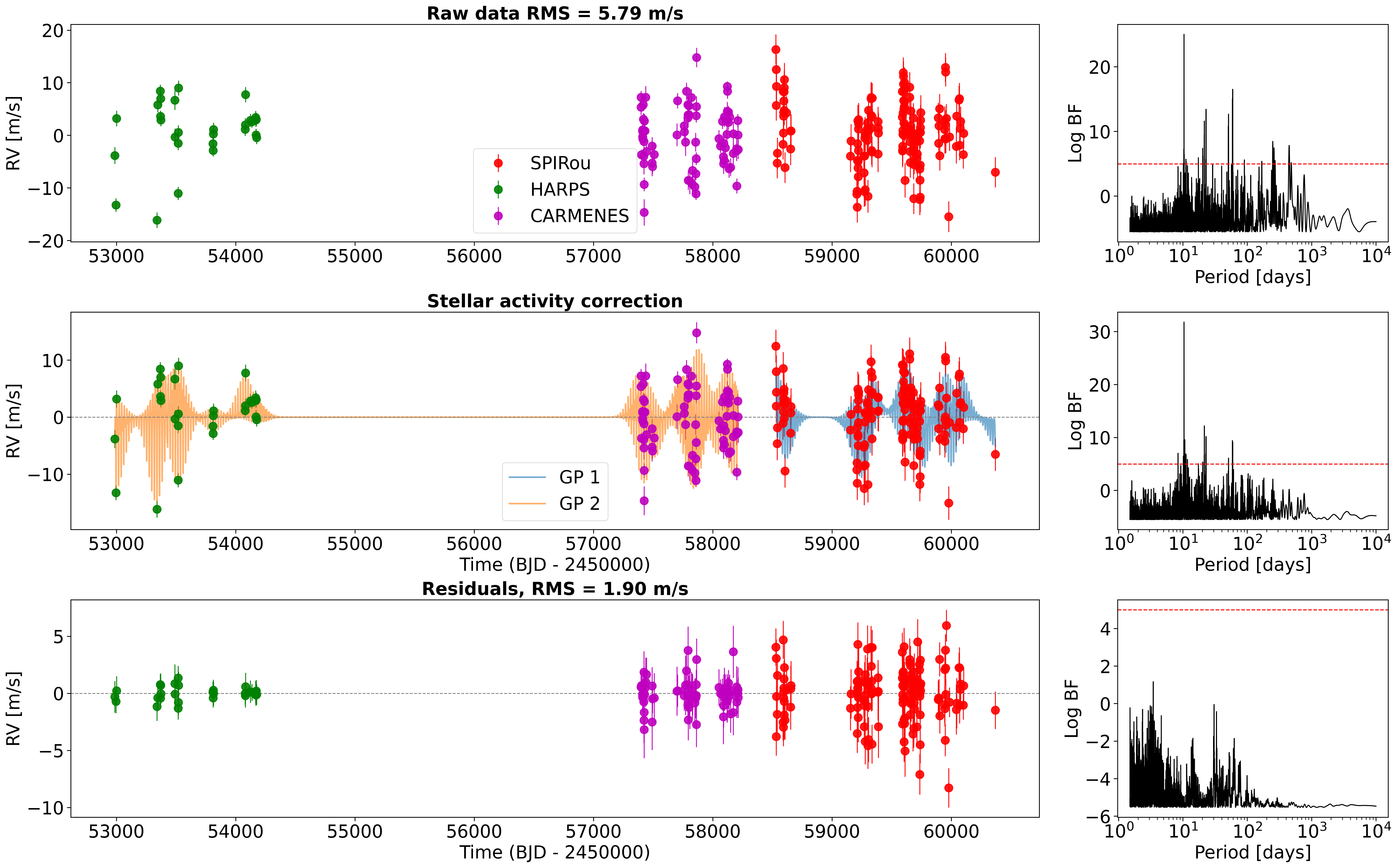} % Reduce size to fit columns
    \caption{Final RV model for Gl\,382. The top panel shows the raw RV data with offsets removed, plotted for SPIRou, HARPS, and CARMENES instruments (RMS = 5.79\,m.s$^{-1}$). The second row illustrates the correction for stellar activity using GPs, with individual GP contributions shown for each grouped instrument. The bottom panel presents the residuals after removing all modeled signals (RMS = 1.90\,m.s$^{-1}$). The rightmost column contains log BF periodograms highlighting significant periods in the RV data, the significance level of 5 is indicated by the horizontal red dashed line.}
    \label{fig:gl382_spirou_RV_analysis}
\end{figure*}

\begin{table*}
\centering
\caption{Orbital and derived planetary parameters from the analysis of Gl\,480.}
\small % Reduce font size
\begin{tabular}{ccccc}
\hline
& SPIRou & Optical & ALL & Priors \\ [0.5ex]
\hline
\noalign{\smallskip}
\multicolumn{5}{c}{\textbf{Instrumental Parameters}} \\
\noalign{\smallskip}
\hline
\noalign{\smallskip}
$\gamma_{SPIRou}$ (m\,s$^{-1}$) & $7915.6^{+1.3}_{-1.3}$ & - &  $7915.6^{+1.2}_{-1.3}$ & $\mathcal{U}\left(-10^{10}, 10^{10}\right)$ \\ 
$\gamma_{CARMENES}$ (m\,s$^{-1}$) & - & $-0.03^{+1.2}_{-1.2}$ & $0.02^{+1.2}_{-1.2}$ & $\mathcal{U}\left(-10^{10}, 10^{10}\right)$ \\ 
$\gamma_{HARPS}$ (m\,s$^{-1}$) & - & $-0.88^{+1.4}_{-1.4}$ & $-0.08^{+1.5}_{-1.5}$ & $\mathcal{U}\left(-10^{10}, 10^{10}\right)$ \\ 
$\sigma_{SPIRou}$ (m\,s$^{-1}$) & $2.4^{+0.4}_{-0.3}$ & - & $2.4^{+0.4}_{-0.3}$ & $\mathcal{MLU}\left(y_\sigma, 2y_{ptp}\right)$ \\ 
$\sigma_{CARMENES}$ (m\,s$^{-1}$) & - &$0.2^{+1.2}_{-0.2}$ & $0.3^{+1.4}_{-0.3}$ & $\mathcal{MLU}\left(y_\sigma, 2y_{ptp}\right)$ \\ 
$\sigma_{HARPS}$ (m\,s$^{-1}$) & - & $0.6^{+1.8}_{-0.6}$ & $0.8^{+2.0}_{-0.8}$ & $\mathcal{MLU}\left(y_\sigma, 2y_{ptp}\right)$ \\ 
\noalign{\smallskip}
\hline
\noalign{\smallskip}
\multicolumn{5}{c}{\textbf{Wapiti Systematics}} \\
\noalign{\smallskip}
\hline
\noalign{\smallskip}
$a_1$ (m/s) & $4.8^{+7.2}_{-7.2}$ & - & $5.0^{+7.2}_{-7.6}$ & $\mathcal{U}\left(-10^{2},10^2\right)$ \\ 
$a_2$ (m/s) & $-11.1^{+5.8}_{-5.8}$ & - & $-11.0^{+5.7}_{-5.7}$ & $\mathcal{U}\left(-10^{2},10^2\right)$ \\ 
$a_3$ (m/s) & $-21.6^{+6.0}_{-6.2}$ & -  & $-21.7^{+6.2}_{-6.0}$ & $\mathcal{U}\left(-10^{2},10^2\right)$  \\ 
$a_4$ (m/s) & $9.2^{+5.0}_{-5.0}$ & -  & $9.1^{+5.0}_{-5.0}$ & $\mathcal{U}\left(-10^{2},10^2\right)$  \\ 
\noalign{\smallskip}
\hline
\noalign{\smallskip}
\multicolumn{5}{c}{\textbf{GP Parameters}} \\
\noalign{\smallskip}
\hline
\noalign{\smallskip}
$A_{SPIRou}$ (m\,s$^{-1}$) & $5.3^{+0.9}_{-0.8}$ & - & $5.2^{+0.9}_{-0.8}$ & $\mathcal{MLU}\left(y_\sigma, 2y_{ptp}\right)$ \\
$\lambda_{SPIRou}$ (days) & $64.5^{+14.9}_{-11.6}$ & - & $64.2^{+15.6}_{-11.5}$ & Posterior$_{\mathrm{dET}}$ \\ 
$\Gamma_{SPIRou}$  & $3.0^{+0.7}_{-0.6}$ & - & $3.0^{+0.7}_{-0.5}$ & Posterior$_{\mathrm{dET}}$ \\
$P_{rot, SPIRou}$ (days) & $20.8^{+0.2}_{-0.2}$ & - & $20.8^{+0.2}_{-0.2}$ & $\mathcal{U}\left(1, 10^{3}\right)$ \\
$A_{Optical}$ (m\,s$^{-1}$) & - & $5.5^{+0.6}_{-0.5}$ & $5.5^{+0.7}_{-0.6}$ & $\mathcal{MLU}\left(y_\sigma, 2y_{ptp}\right)$ \\
$\lambda_{Optical}$ (days) & - & $71.0^{+19.0}_{-13.5}$ & $72.1^{+21.3}_{-14.5}$ & $\mathcal{LU}\left(\delta t_{av}, 10t_{span}\right)$ \\ 
$\Gamma_{Optical}$  & - & $13.2^{+5.2}_{-5.4}$ & $12.6^{+5.3}_{-6.4}$ & $\mathcal{LU}\left(.1, 50\right)$ \\
$P_{rot, Optical}$ (days) & - & $21.5^{+0.1}_{-0.1}$ & $21.5^{+0.2}_{-0.1}$ & $\mathcal{U}\left(1, 10^{3}\right)$ \\ 
\noalign{\smallskip}
\hline
\noalign{\smallskip}
\end{tabular}
\begin{minipage}{\textwidth} % This ensures the note is centered and spans the table width
\small
The parameters derived from the \texttt{Wapiti} corrected RV data from SPIRou are compared to those obtained by combining optical data from HARPS and CARMENES. The parameters derived by combining all datasets correspond to the ALL column. 
\end{minipage}
\label{gl382_parameters}
\end{table*}

\section{Conclusion}\label{sec:conclusion}

{In this study, we performed a detailed RV analysis of the nearby M dwarf stars Gl\,480 and Gl\,382 by combining nIR data from the SPIRou spectropolarimeter with non-simultaneous optical data from the HARPS and CARMENES spectrographs. Using advanced correction techniques such as the LBL method and the Wapiti algorithm, we effectively reduced instrumental noise and stellar activity effects, improving the detection of planetary signals.}
\par
{For Gl\,480, our analysis confirms the presence of a potential planet with an orbital period of $9.5537 \pm 0.0005$\,d and a minimum mass of $8.8 \pm 0.7$\,M$_\oplus$.  This result provides a more precise mass estimate and differs significantly from the previously reported value of $13.2 \pm 1.7$\,M$_\oplus$. A 17.8\,d signal also appeared in the nIR data but was linked to incomplete correction of stellar activity. This signal, observed in the dET and dLW activity indicators, was suppressed only when the GP models used priors based on stellar activity.}

\par

{Additionally, a low-amplitude signal with a period of 6.4\,d was detected when combining all datasets. However, the statistical significance of this signal depends critically on the choice of prior distributions in our quasi-periodic GP models. Specifically, this signal reached detection significance only when the priors were informed by the dLW stellar activity indicator. In contrast, when using non-informative \citep{Camacho2023} or dET-based priors, the signal did not surpass the detection threshold. We also studied the non-detection of this signal in the optical using an injection-recovery test and found that in 81\% of our simulations, this signal could not be detected with these data. Further high-precision RV observations would be necessary to confirm or refute the planetary nature of this candidate signal and to better constrain its orbital parameters., this signal could not be detected with these data. If confirmed, this planet would be in a near 3:2 resonance with the 9.5\,d planet and its minimum mass would be of $2.7 \pm 0.6$\,M$_\oplus$.}
\par
{For Gl\,382, despite employing the same rigorous methodology, we did not find any evidence for planetary companions. The RV variations in this system are dominated by stellar activity. Based on our combined dataset, we estimated that our detection limit for a potential companion orbiting Gl\,382 within its HZ corresponds to an upper mass limit of  M$_\mathbf{upper} = 5.9 \pm 0.8$\,M$_\oplus$.}
\par
{The stellar activity in Gl\,382 is particularly intriguing due to the distinct ways it manifests in the optical and nIR domains: while the optical data revealed both the rotational period and its harmonic, the nIR data predominantly exhibited the first harmonic as well as a 51\,d signal (which was present in both SPIRou and HARPS proxies). This wavelength-dependent behavior of activity-induced signals highlights the complexity of magnetic phenomena in M dwarfs and suggests that Gl\,382 could serve as a promising laboratory for studying stellar magnetic activity and refining activity correction methods in RV analyses.}
\par
{The use of GP regression models with varying prior distributions played a critical role in our analysis. The priors were carefully constrained using stellar activity indicators such as the dET and dLW time series, which provided insights into the stellar rotational periods and activity timescales. This approach was crucial in identifying and mitigating activity-related signals in nIR RVs, such as the 17.8\,d signal in Gl\,480 and the 51.1\,d signal in Gl\,382, highlighting the importance of integrating activity diagnostics into RV analyses.}
\par 
{In contrast, the set of uninformative priors from \citet{Camacho2023} was sufficient to correct for stellar activity in the optical datasets. A similar difference in activity correction between optical and nIR data was also reported by \citet{pia2023}, who suggested that this could indicate a different underlying process driving RV variations in the nIR, possibly linked to the Zeeman effect.}
\par 
{An alternative approach for incorporating such diagnostics would have been to use a multi-dimensional GP, as introduced by \citet{Rajpaul_2015}. This method enables the simultaneous modeling of stellar activity in RV time series by modeling both the RV data and activity indicators with a GP  $G$ and its time derivative $\dot G$. However, multi-dimensional GPs are not yet supported within the Wapiti framework, and implementing this capability could be the object of future work.}
\par
{Interestingly, for both targets, the GP amplitude is similar in the optical and nIR, a result also reported by \citet{pia2023}. This suggests that, for these two stars, observing in the nIR does not necessarily provide an advantage over the optical, which is expected for this type of M dwarf. As noted by \citet{reiners2010}, the improvement in precision when moving to the nIR is most significant for M4 and later spectral types.}

\par

{Additionally, in the case of the nIR RVs of Gl\,480, stellar activity was not clearly apparent, meaning that we did not observe a strong signal at either the rotational period or its first harmonic, as we did with Gl\,382. However, the presence of the 17.8\,d signal prompted us to model stellar activity, ultimately leading to our best-fit model. Nevertheless, in the absence of a clear activity-related signal, we would still recommend modeling stellar activity rather than disregarding it, as advised by \citet{Stock2023}.}

\par 

{This study demonstrates  the power of combining nIR and optical RV data to detect and characterize exoplanets around active M dwarf stars. The complementary nature of multi-wavelength observations is key to distinguishing true planetary signals from stellar activity. However, detecting low-amplitude signals remains challenging and depends heavily on how stellar activity is modeled.}
\par 
{Looking forward, extended RV monitoring of Gl\,480 with higher cadence and precision will be essential to confirm the tentative 6.4\,d signal and explore the possibility of additional planetary companions. Additionally, the complex and dynamic stellar activity observed in Gl\,382 provides a valuable opportunity to refine techniques for stellar activity correction, with broader implications for improving exoplanet detection across diverse stellar types.}

\begin{acknowledgements}
This study has ben partially supported through the grant EUR TESS N°ANR-18-EURE-0018 in the framework of the Programme des Investissements d'Avenir.
JFD and CM acknowledge funding from the European Research Council (ERC) under the H2020 research \& innovation program (grant agreements \#740651 NewWorlds).  
\par
X.D. and A.C. acknoweldge funding by the French National Research Agency in the framework of the Investissements d’Avenir program (ANR-15-IDEX-02), through the funding of the “Origin of Life” project of the Université Grenoble Alpes.
\par
EA, CC, NJC \& RD acknowledge the financial support of the FRQ-NT through the {\it Centre de recherche en astrophysique du Québec}.
\par
E.M. acknowledges funding from FAPEMIG under project number APQ-02493-22
and research productivity grant number 309829/2022-4 awarded by the CNPq,
Brazil.
\par 
This paper makes use of data from the first public release of the WASP data (Butters et al. 2010) as provided by the WASP consortium and services at the NASA Exoplanet Archive, which is operated by the California Institute of Technology, under contract with the National Aeronautics and Space Administration under the Exoplanet Exploration Program.
\par 
This research made use of the following software tools: \texttt{NumPy}; a fundamental package for scientific computing with Python. \texttt{pandas}; a library providing easy-to-use data structures and data analysis tools. \texttt{Astropy}; a community-developed core Python package for Astronomy. \texttt{SciPy}; a library for scientific computing and technical computing. \texttt{RadVel}; a python package for modeling radial velocity data. \texttt{matplotlib}; a plotting library for the Python programming language. \texttt{tqdm}; a library for creating progress bars in the command line. \texttt{seaborn}; a data visualization library based on matplotlib. \texttt{wpca}; a python package for weighted principal component analysis.  \texttt{emcee}; a python package for MCMC sampling. \texttt{PyAstronomy}; a collection of astronomical tools for Python. \texttt{celerite}; a fast Gaussian process library in Python. \texttt{george}: a module for Gaussian process regression in Python.

\end{acknowledgements}
\clearpage

\bibliographystyle{aa}\bibliography{ref.bib}{}
\begin{appendix}
\onecolumn
\section{Supplementary material regarding the MCMC analysis}

\subsection{Gl\,480 RV analysis}\label{sec:RVanalysis480}

In Figures \ref{fig:GL480_det_activity} and \ref{fig:GL480_dlw_activity}, we present the time series of the stellar activity indicators dET and dLW, respectively, along with their fitted GPs. 

\begin{figure*}[h!]
    \centering
    \includegraphics[width=.7\linewidth]{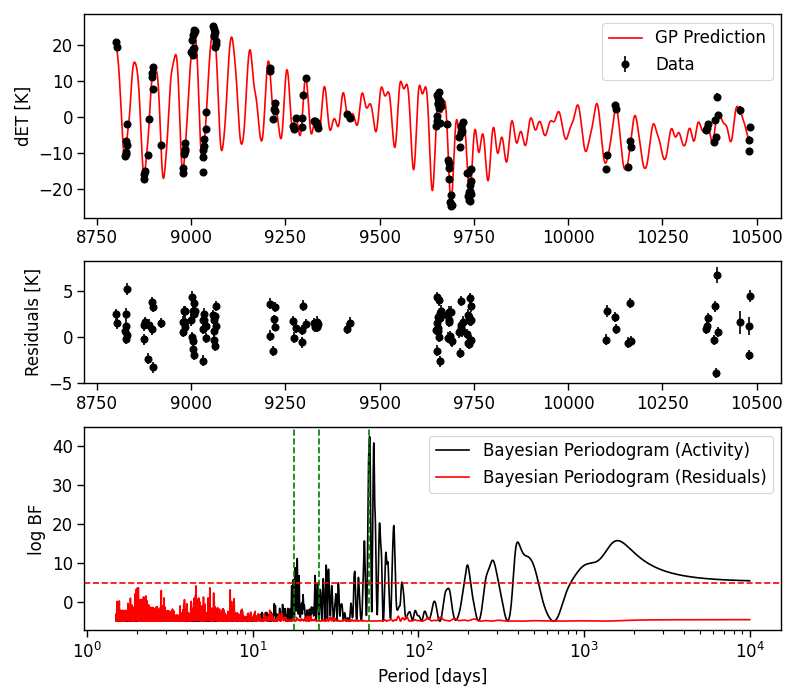}
    \caption{Top panel shows the dET time series for GL\,480, and the fitted GP model overlaid in red. The middle panel displays the residuals after subtracting the GP model from the observations. The bottom panel presents the Bayesian periodograms for the activity signal (black curve) and residuals (red curve), plotted as a function of period (in days). The dashed green lines in the bottom panel mark periodicities at the stellar rotation period (P$_{\mathbf{rot}}$), half rotation period (P$_{\mathbf{rot}}$/2), and one-third rotation period (P$_{\mathbf{rot}}$/3).  The significance level of log BF $= 5$ is indicated by the horizontal red dashed line.}
    \label{fig:GL480_det_activity}
\end{figure*}

\begin{figure*}[h!]
    \centering
    \includegraphics[width=.7\linewidth]{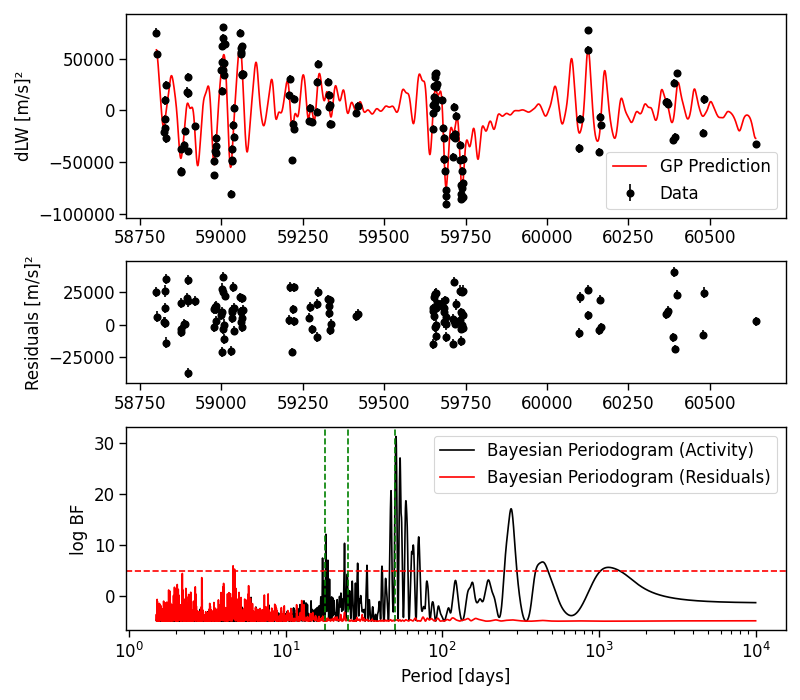}
    \caption{Top panel shows the dLW time series for GL\,480, and the fitted GP model overlaid in red. The middle panel displays the residuals after subtracting the GP model from the observations. The bottom panel presents the Bayesian periodograms for the activity signal (black curve) and residuals (red curve), plotted as a function of period (in days). The dashed green lines in the bottom panel mark periodicities at the stellar rotation period (P$_{\mathbf{rot}}$), half rotation period (P$_{\mathbf{rot}}$/2), and one-third rotation period (P$_{\mathbf{rot}}$/3).  The significance level of log BF $= 5$ is indicated by the horizontal red dashed line.}
    \label{fig:GL480_dlw_activity}
\end{figure*}

The corresponding corner plots, derived from the MCMC analysis of the dET and dLW time series, are shown in Figures \ref{fig:gl480_dtemp_mcmc} and \ref{fig:GL480_dlw_mcmc}.

\begin{figure*}[h!]
    \centering
    \includegraphics[width=.8\linewidth]{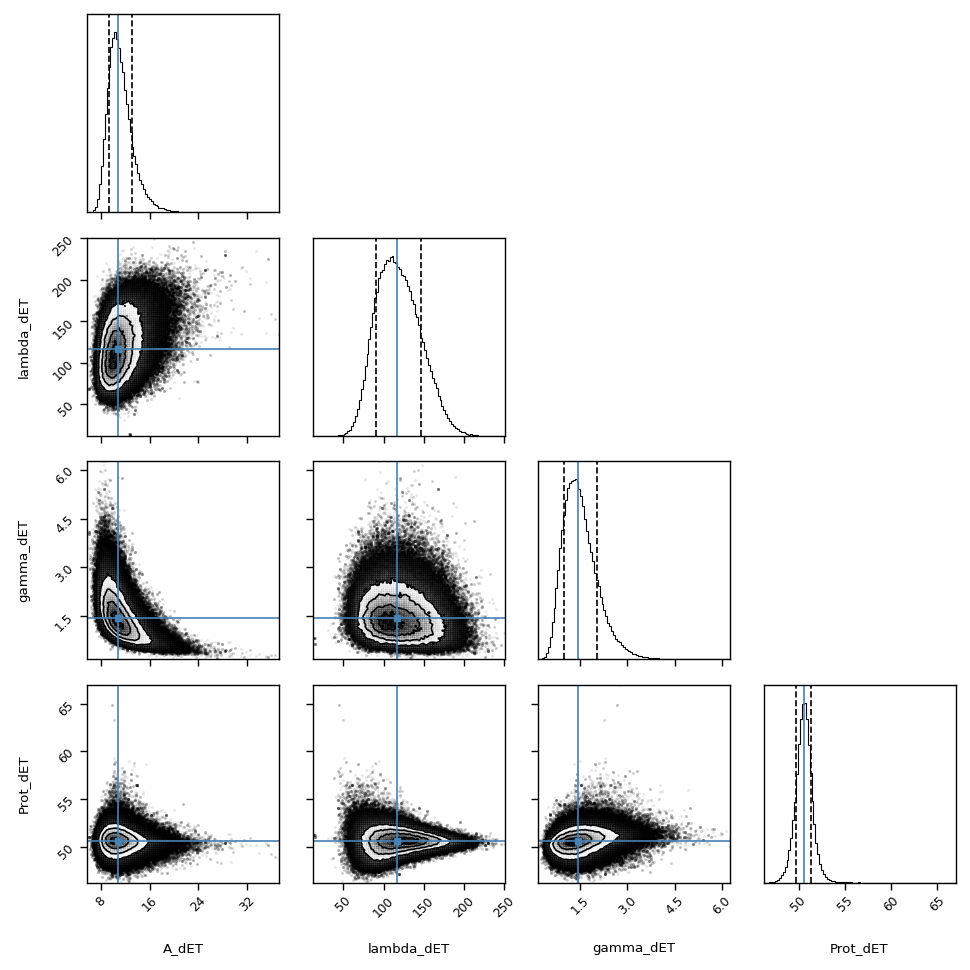}
    \caption{Corner plot displaying the results of MCMC analysis on the dET time series of Gl\,480}
    \label{fig:gl480_dtemp_mcmc}
\end{figure*}

\begin{figure*}[h!]
    \centering
    \includegraphics[width=.8\linewidth]{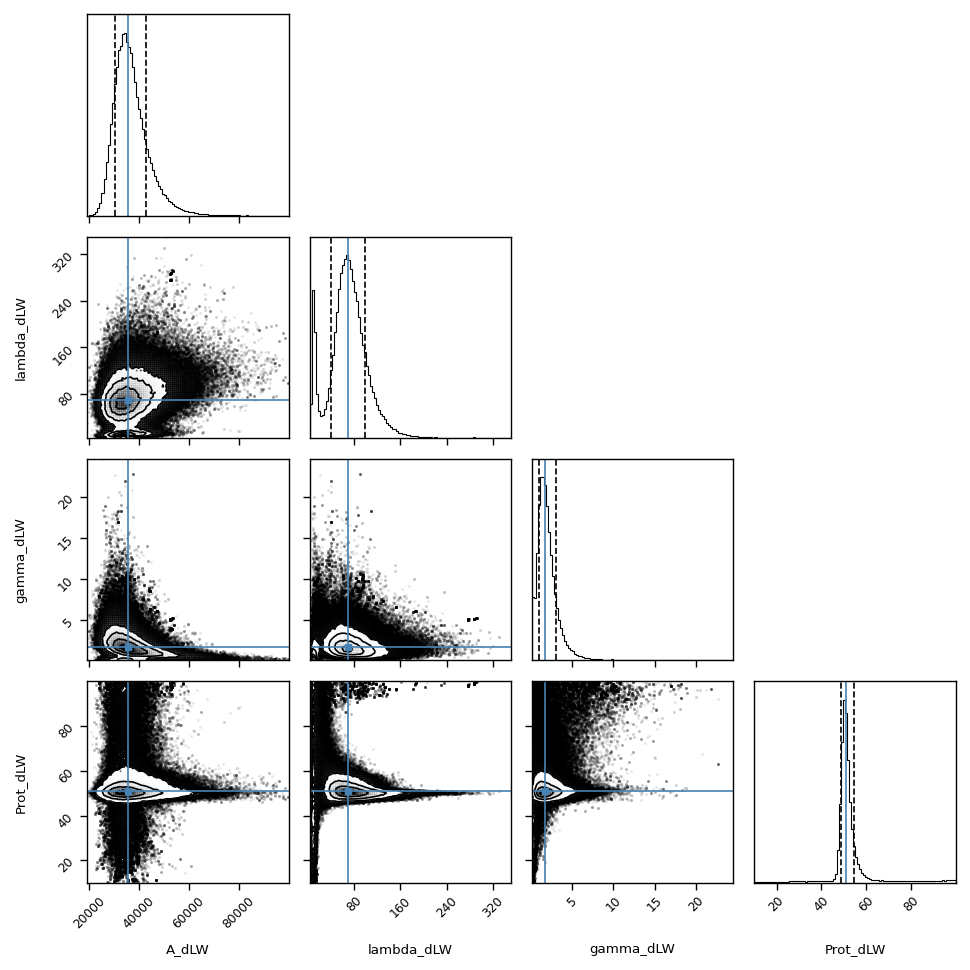}
    \caption{Corner plot displaying the results of MCMC analysis on the dLW time series of Gl\,480}
    \label{fig:GL480_dlw_mcmc}
\end{figure*}

Figure \ref{fig:GL480_distributions_comparison} compares the posterior distributions of the semi-amplitude K and the orbital period P of Gl\,480 across different datasets. Additionally, the posterior distribution of the rotational period retrieved from the SPIRou dataset is shown, using the posterior distribution of the same hyperparameters from the dLW analysis as a prior. This distribution is compared to those derived from the activity indicators dLW and dET, highlighting the consistency and differences among the datasets.

\begin{figure*}[h!]
    \centering
    \includegraphics[width=\linewidth]{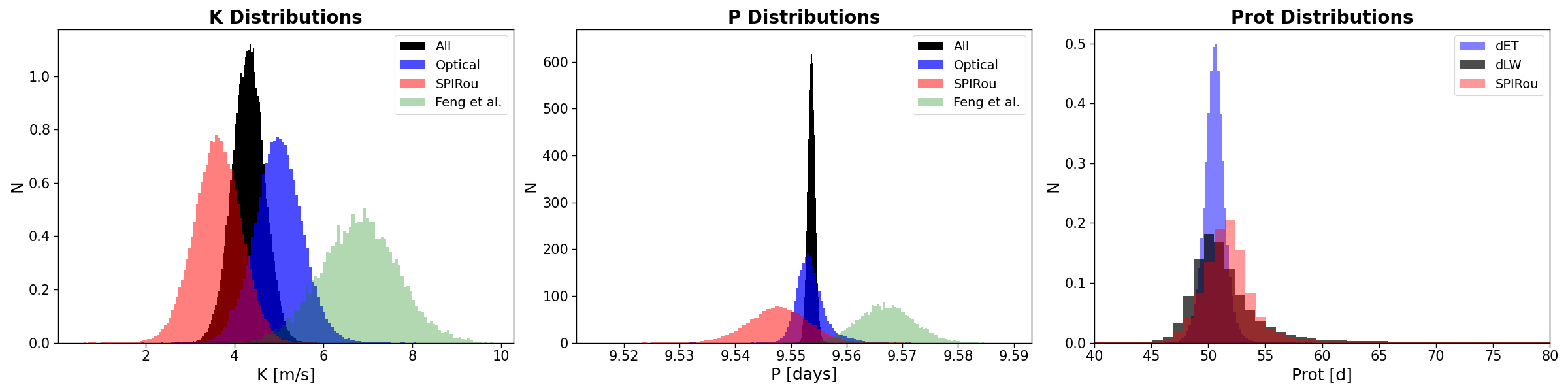}
    \caption{Posterior distributions of the key parameters derived from the MCMC analysis for GL\,480. The left panel shows the posterior distribution of the semi-amplitude K for the combined dataset (black), optical data (blue), SPIRou data (red), and the reference distribution from \citet{2020ApJS..250...29F}. The middle panel presents the posterior distributions of the orbital period P, using the same color scheme. The right panel displays the posterior distributions of the stellar rotation period $\mathrm{P_{rot}}$ derived from the dET (blue) and dLW (black) time series, and the SPIRou dataset (red). All histograms are normalized to unit area for comparison.}
    \label{fig:GL480_distributions_comparison}
\end{figure*}

The final corner plot from the MCMC analysis of the RV model is divided for clarity across Figures \ref{fig:GL480_instrumental}, \ref{fig:GL480_planets}, and \ref{fig:GL480_mcmc_activity}. These figures respectively display the posterior distributions of the instrumental and Wapiti parameters, the orbital parameters of the two planets, and the hyperparameters for the two GPs applied to the SPIRou and optical datasets.

\begin{figure*}[h!]
    \centering
    \includegraphics[width=.8\linewidth]{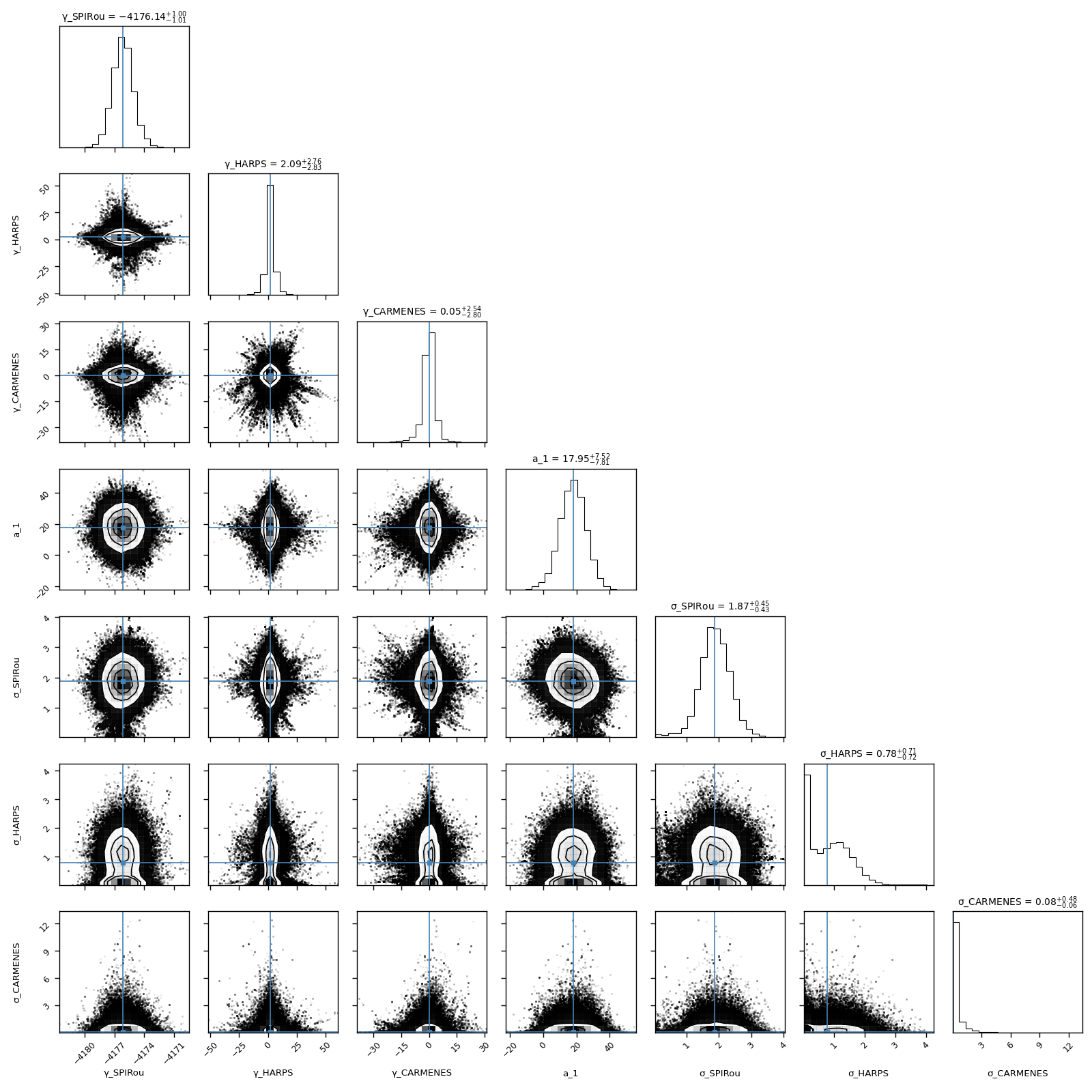}
    \caption{Corner plot displaying the results for the offsets, the Wapiti systematics and the jitters of the MCMC analysis on the RV time series of Gl\,480}
    \label{fig:GL480_instrumental}
\end{figure*}

\begin{figure*}[h!]
    \centering
    \includegraphics[width=.8\linewidth]{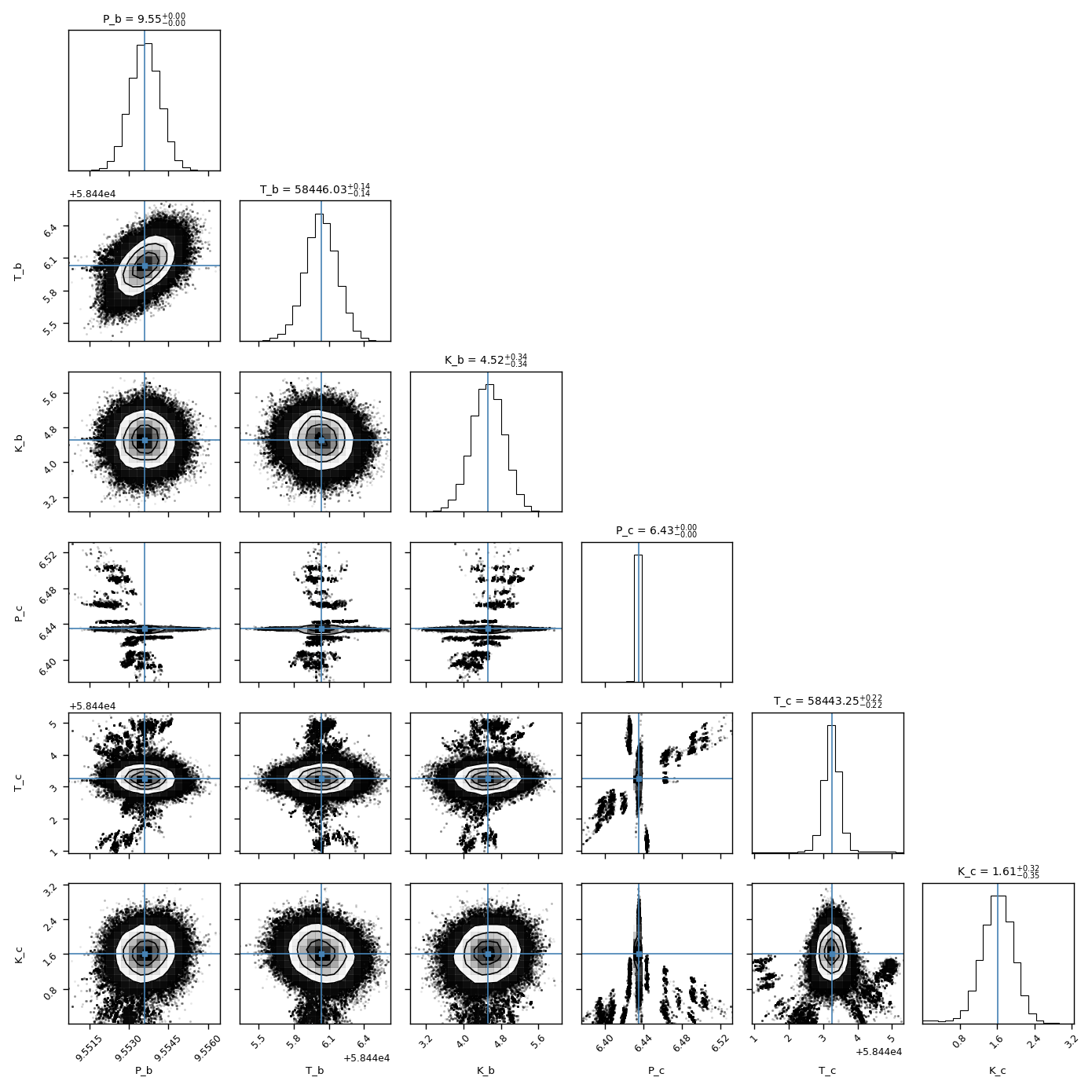}
    \caption{Corner plot displaying the results for the orbital parameters of the MCMC analysis on the RV time series of Gl\,480}
    \label{fig:GL480_planets}
\end{figure*}

\begin{figure*}[h!]
    \centering
    \includegraphics[width=.8\linewidth]{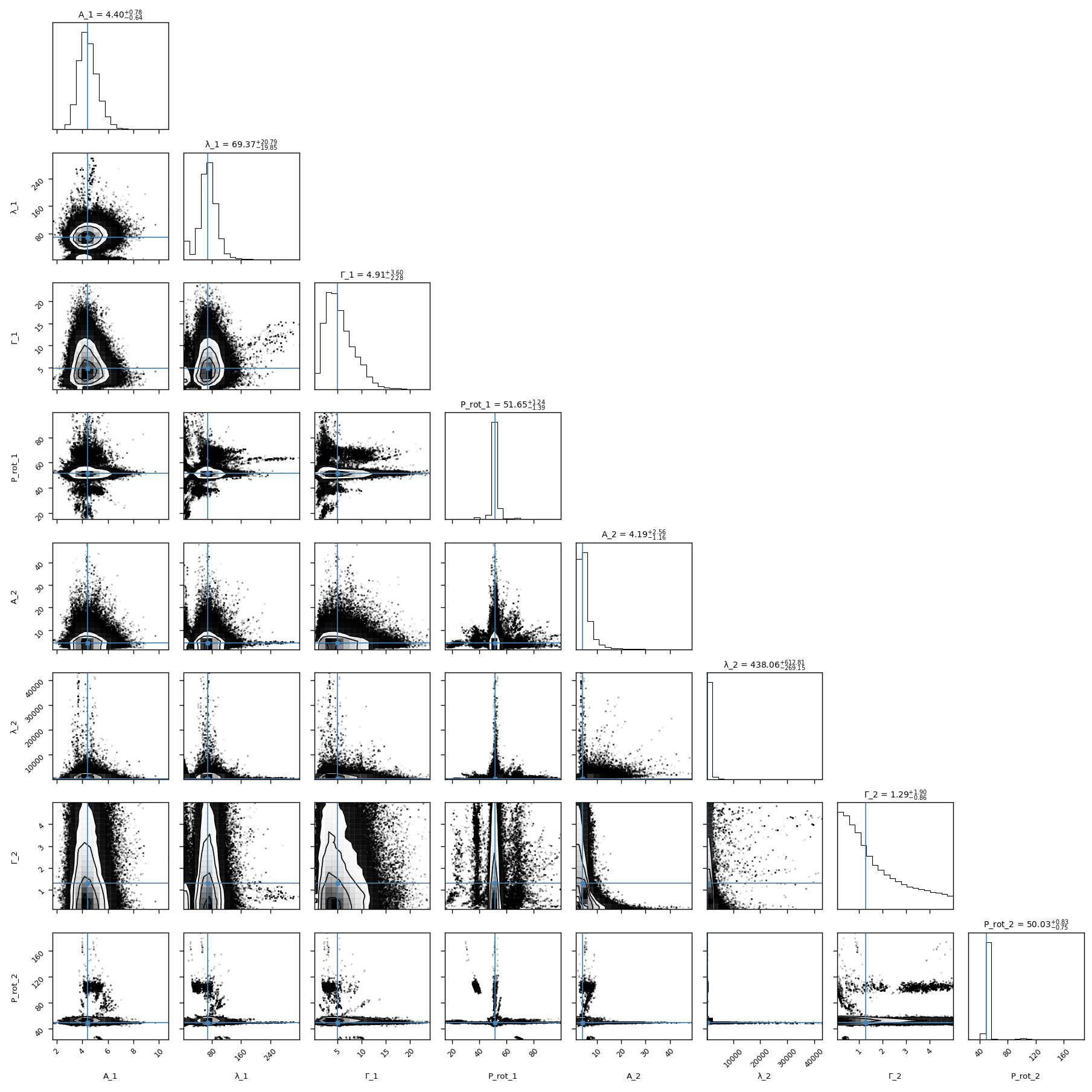}
    \caption{Corner plot displaying the results for the offsets, the Wapiti systematics and the jitters of the MCMC analysis on the RV time series of Gl\,480}
    \label{fig:GL480_mcmc_activity}
\end{figure*}

\clearpage

\subsection{Gl\,382 RV analysis}\label{sec:RVanalysis382}

In Figures \ref{fig:GL382_det_activity}, we present the time series of the stellar activity indicators dET along with its fitted GPs. The corresponding corner plots, derived from the MCMC analysis of the dET and dLW time series, are shown in Figures \ref{fig:gl382_dtemp_mcmc}.

\par

The final corner plot from the MCMC analysis of the RV model is divided for clarity across Figures \ref{fig:GL382_instrumental}, and \ref{fig:GL382_mcmc_activity}. These figures respectively display the posterior distributions of the instrumental and Wapiti parameter, and the hyperparameters for the two GPs applied to the SPIRou and optical datasets.
\begin{figure*}[h!]
    \centering
    \includegraphics[width=.7\linewidth]{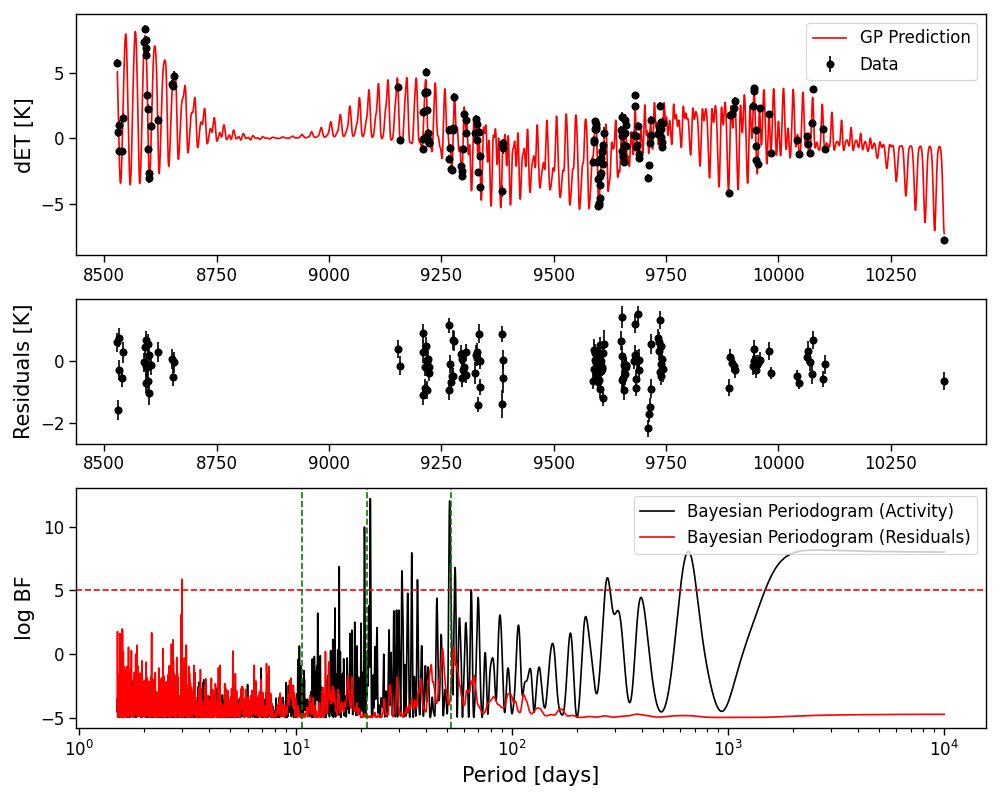}
    \caption{Top panel shows the dET time series for GL\,382, and the fitted GP model overlaid in red. The middle panel displays the residuals after subtracting the GP model from the observations. The bottom panel presents the Bayesian periodograms for the activity signal (black curve) and residuals (red curve), plotted as a function of period (in days). The dashed green lines in the bottom panel mark periodicities at the stellar rotation period (P$_{\mathbf{rot}}$), half rotation period (P$_{\mathbf{rot}}$/2), and the 51\,d period value.  The significance level of log BF $= 5$ is indicated by the horizontal red dashed line.}
    \label{fig:GL382_det_activity}
\end{figure*}

\begin{figure*}
    \centering
    \includegraphics[width=.8\linewidth]{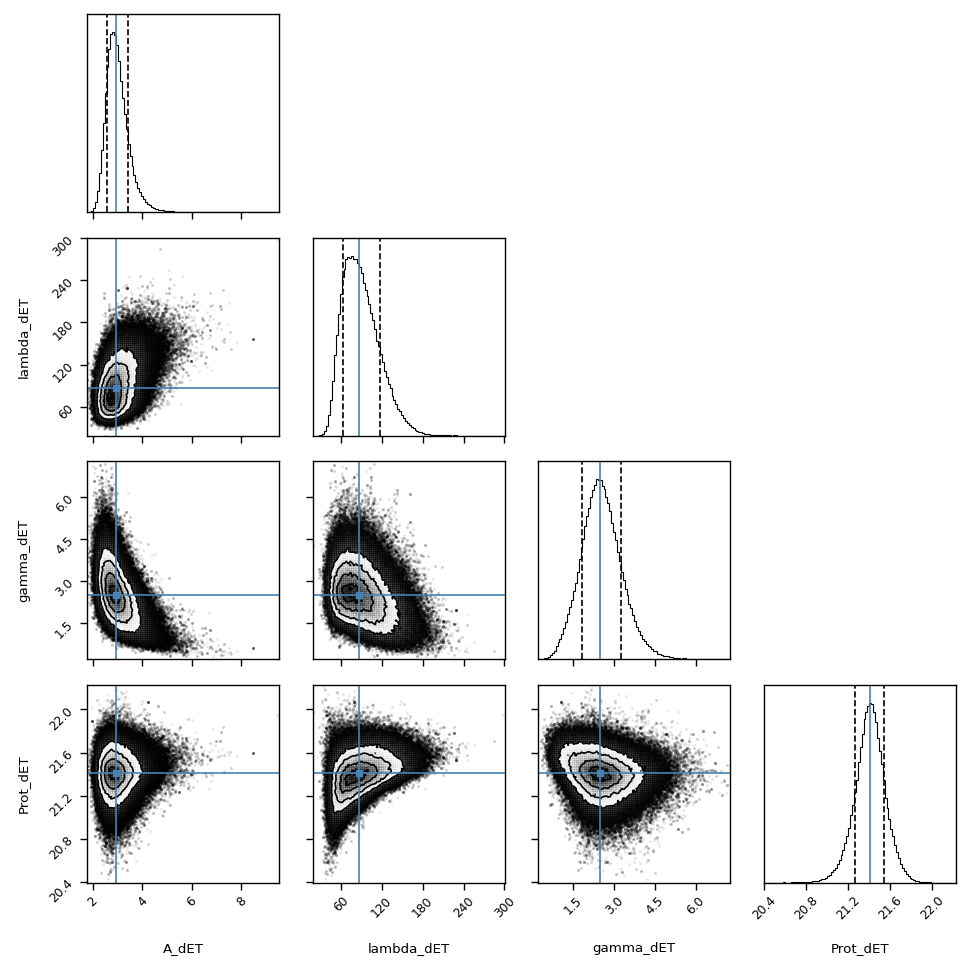}
    \caption{Corner plot displaying the results of MCMC analysis on the dET time series of Gl\,382}
    \label{fig:gl382_dtemp_mcmc}
\end{figure*}

\begin{figure*}
    \centering
    \includegraphics[width=.8\linewidth]{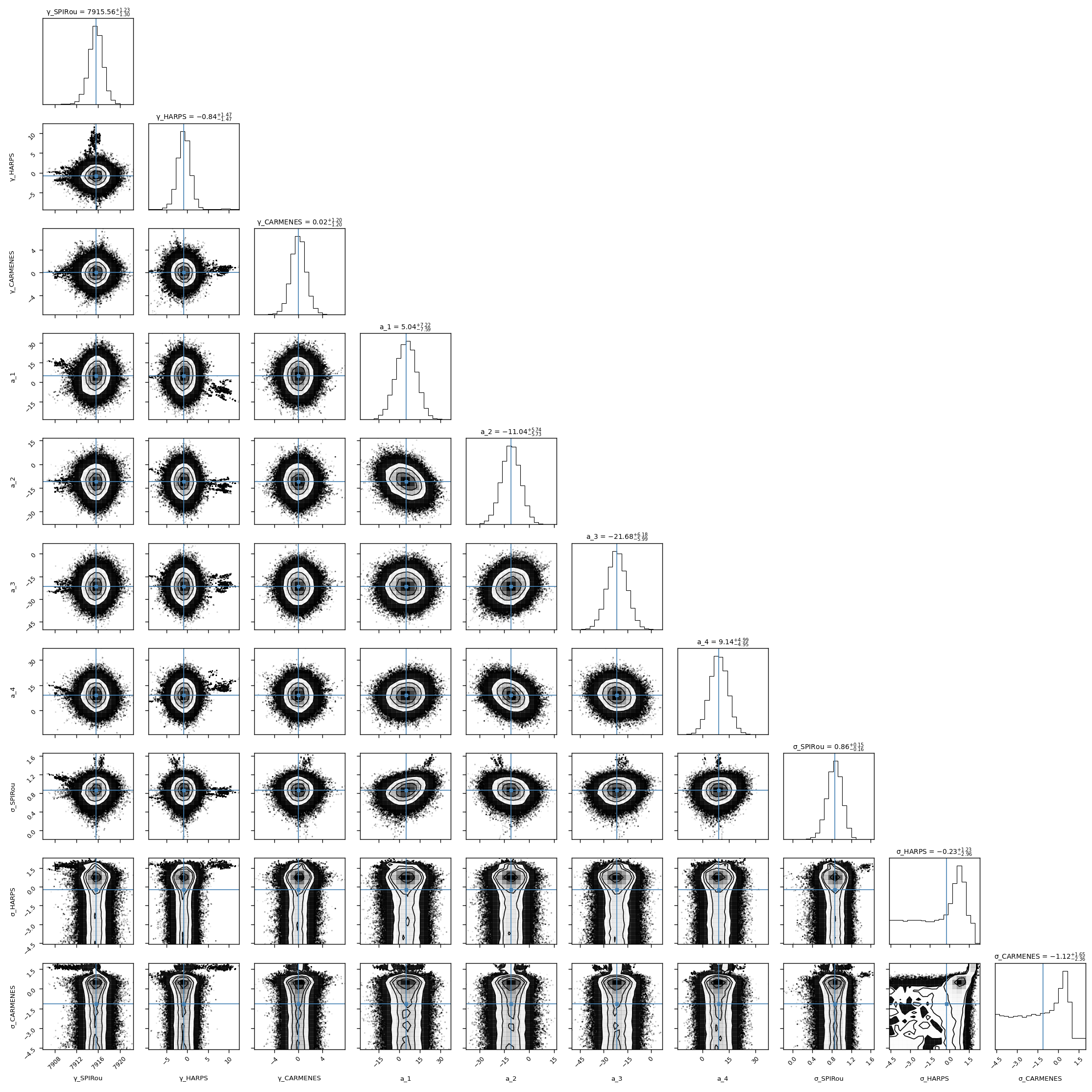}
    \caption{Corner plot displaying the results for the offsets, the Wapiti systematics and the jitters of the MCMC analysis on the RV time series of Gl\,382}
    \label{fig:GL382_instrumental}
\end{figure*}

\begin{figure*}
    \centering
    \includegraphics[width=\linewidth]{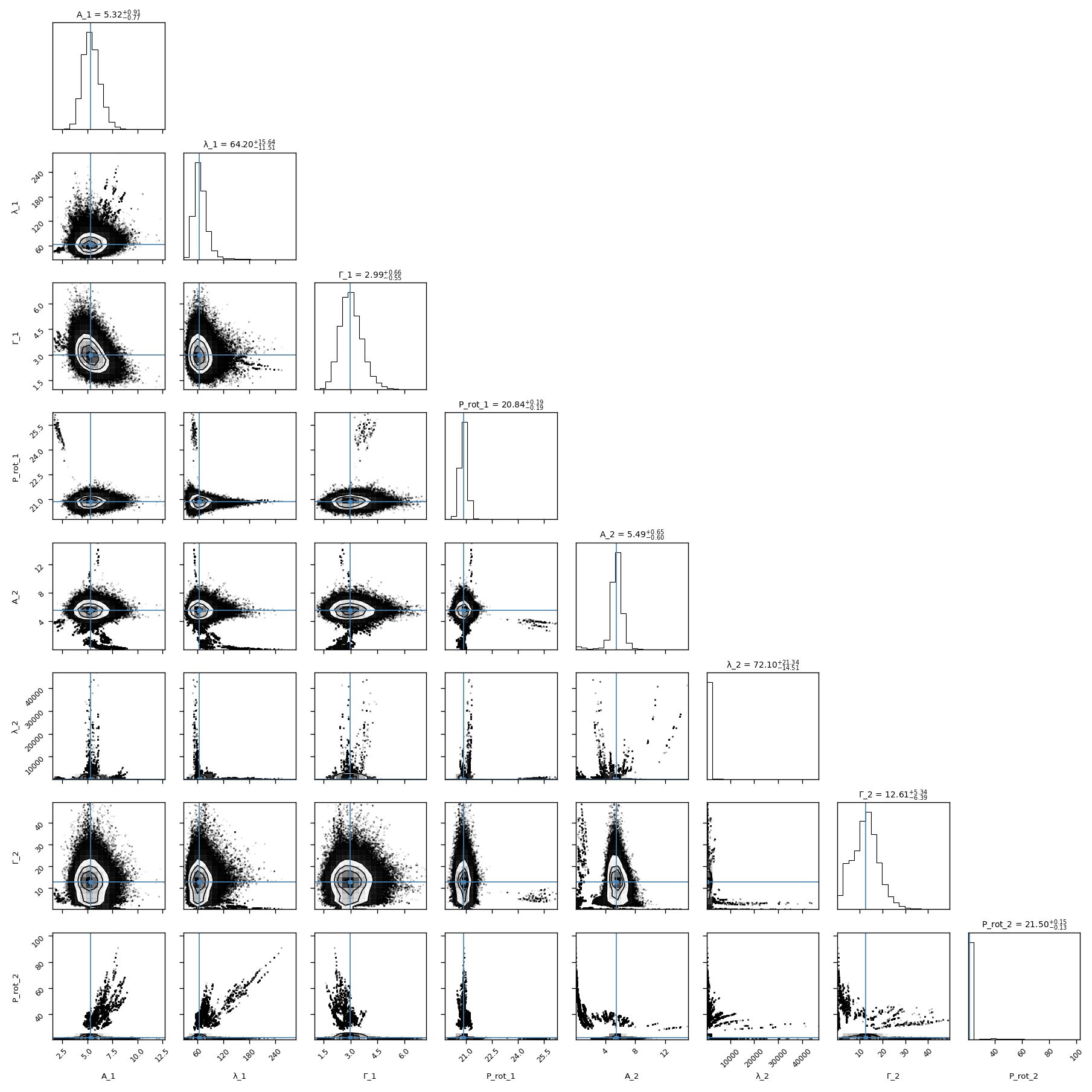}
    \caption{Corner plot displaying the results for the GP hyperparameters applied on the SPIRou and optical data of the MCMC analysis on the RV time series of Gl\,382}
    \label{fig:GL382_mcmc_activity}
\end{figure*}

\end{appendix}
\end{document}